\documentclass[12pt]{article}

\usepackage{graphicx}

\newcommand{\nc}{\newcommand}
\nc{\rnc}{\renewcommand}
\rnc{\theequation}{\thesection.\arabic{equation}}
\rnc{\arraystretch}{1.3}

\def\nsection#1{\setcounter{equation}{0}\section{#1}}

\font\tenmsb=msbm10 scaled \magstep 1
\font\sevenmsb=msbm7 scaled \magstep1
\font\fivemsb=msbm5 scaled \magstep1
\newfam\msbfam
\textfont\msbfam=\tenmsb
\scriptfont\msbfam=\sevenmsb
\scriptscriptfont\msbfam=\fivemsb
\def\Bbb#1{{\fam\msbfam\relax#1}}

\def\os{m}
\def\Re{\mathop{\rm Re}}
\def\Im{\mathop{\rm Im}}
\def\cosec{\mathop{\rm cosec}}
\def\Mult#1#2{\left[{#1 \atop #2}\right]}
\def\m{\hspace{-0.15mm} - \hspace{-0.25mm} }

\def\e{\mbox{e}}
\def\i{\mbox{i}}
\def\smbox#1{\mbox{\scriptsize #1}}

\def\case#1#2{{\textstyle{#1\over #2}}}
\def\sc{\scriptstyle}

\def\W#1#2#3#4#5{W\!\left(\left.\matrix{#1&#2\cr #4&#5}\right|#3\right)}

\def\BLt#1#2#3#4{\renewcommand{\arraystretch}{0.2}
                    B\!\left(\!\!\begin{array}{rr|c}
                    #3 & & \\ & #2 & #4 \\ #1 & & \end{array}\!\!\!\right)}
\def\BRt#1#2#3#4{\renewcommand{\arraystretch}{0.2}
                    B\!\left(\!\!\!\begin{array}{rr|c}
                    & #3 & \\ #2 & & #4\\ & #1 & \end{array}\!\!\right)}

\def\eqref#1{(\ref{#1})}
\def\fb{f_{\rm bulk}}

\def\fs{f_{\rm bdy}}
\def\T{\mbox{\boldmath $T$}}
\def\F{\mbox{\boldmath $F$}}
\def\R{\mbox{\boldmath $R$}}
\def\C{\mbox{\boldmath $C$}}
\def\I{\mbox{\boldmath $I$}}
\def\x{\mbox{\boldmath $x$}}

\def\z{\mbox{\boldmath $z$}}

\def\a{\mbox{\boldmath $a$}}

\def\smb#1{\mbox{\scriptsize\boldmath $#1$}}

\def\b{\mbox{\boldmath $b$}}
\def\c{\mbox{\boldmath $c$}}
\def\D{\mbox{\boldmath $D$}}

\def\V{\mbox{\boldmath $V$}}
\def\cA{\mbox{$\mathcal{A}$}}

\def\td#1{\mbox{$\tilde{#1}$}}
\def\tD{\mbox{$\td\D$}}
\def\md{\circle*{0.03}}
\def\sCyl{{\mbox{\scriptsize Cyl}}}
\def\sTorus{{\mbox{\scriptsize Torus}}}
\def\sKlein{{\mbox{\scriptsize Klein}}}
\def\sMobius{{\mbox{\scriptsize M\"obius}}}
\def\mub{\bar{\mu}}

\nc{\beq}{\begin{equation}}
\nc{\beqa}{\begin{eqnarray}}
\nc{\eql}[1]{\label{Eqn#1}}
\nc{\bleq}[1]{\beq\eql{#1}}
\nc{\eeq}{\end{equation}}
\nc{\eeqa}{\end{eqnarray}}
\nc{\noeqno}{\nonumber\\}
\nc{\sm}[1]{{\scriptstyle #1}}
\nc{\ssm}[1]{{\scriptscriptstyle #1}}
\nc{\DD}{{\cal D}}
\rnc{\L}{{\cal L}}
\nc{\Wpic}[5]{W\!\left(\,\begin{array}{@{}cc|}#4&#3\\#1&#2\end{array}
\;#5\right)}
\nc{\Wf}[6]{W^{#1}\!\!\left(\,\begin{array}{@{}cc|}#5&#4\\#2&#3
\end{array}\;#6\right)}
\nc{\BL}[4]{B_{\mbox{\tiny L}}\!\!\left(\left.\!\!
\begin{array}{c}#3\\#1\end{array}\,#2\:\right|#4\right)}
\nc{\BR}[4]{B_{\mbox{\tiny R}}\!\!\left(\left.\!#2\,
\begin{array}{c}#3\\#1\end{array}\!\right|#4\right)}
\nc{\BtildeL}[4]{\tilde{B}_{\mbox{\tiny L}}\!\!\left(\left.\!\!
\begin{array}{c}#3\\#1\end{array}\,#2\:\right|#4\right)}
\nc{\BtildeR}[4]{\tilde{B}_{\mbox{\tiny R}}\!\!\left(
\left.\!#2\,\begin{array}{c}#3\\#1\end{array}\!\right|#4\right)}
\nc{\BLf}[5]{B^{#1}_{\mbox{\tiny L}}\!\!\left(\left.\!\!
\begin{array}{c}#4\\#2\end{array}\,#3\:\right|#5\right)}
\nc{\BRf}[5]{B^{#1}_{\mbox{\tiny R}}\!\!\left(\left.\!#3\,
\begin{array}{c}#4\\#2\end{array}\!\right|#5\right)}
\nc{\xiL}{\xi_{\mbox{\tiny L}}\!}
\nc{\xiR}{\xi_{\mbox{\tiny R}}\!}
\nc{\aL}{a_{\mbox{\tiny L}}}
\nc{\aR}{a_{\mbox{\tiny R}}}
\rnc{\l}{\lambda}
\nc{\lm}{{\textstyle\frac{\l-\m}{2}}}
\rnc{\t}{\theta}
\nc{\N}{{\scriptscriptstyle N}}
\rnc{\vec}[1]{\mbox{\boldmath$#1$}}
\renewcommand{\ss}{\scriptstyle}

\newcommand{\pp}[2]{\makebox(0,0)[#1]{$\ss#2$}}
\nc{\pos}[2]{\makebox(0,0)[#1]{$#2$}}
\nc{\spos}[2]{\makebox(0,0)[#1]{$\sm{#2}$}}
\nc{\text}[6]{\begin{picture}(#1,#2)
\put(#3,#4){\pos{#5}{\displaystyle#6}}\end{picture}}
\nc{\dl}[3]{\put(#1,#2){\makebox(#3,0){\dotfill}}}
\rnc{\d}[2]{\put(#1,#2){\spos{}{\bullet}}}
\nc{\dd}[3]{\multiput(#1,#2)(0,1){#3}{\spos{}{\bullet}}}
\rnc{\u}{\begin{picture}(0,0)
\put(-0.23,0){\spos{r}{u}}\end{picture}}
\nc{\uq}{\begin{picture}(0,0)
\put(-0.8,0){\spos{l}{\;u\!+\!q\l}}\end{picture}}
\nc{\uql}{\begin{picture}(0,0)
\put(-0.8,0){\spos{l}{u\!+\!(\!q\!-\!1\!)\l}}\end{picture}}
\nc{\uqll}{\begin{picture}(0,0)
\put(-0.8,0){\spos{l}{u\!+\!(\!q\!-\!2\!)\l}}\end{picture}}
\nc{\uuq}{\begin{picture}(0,0)
\put(-0.06,0){\spos{}{-\!2u\!-\!q\l\!+\!\m}}\end{picture}}
\nc{\uuql}{\begin{picture}(0,0)
\put(-0.1,0.3){\spos{}{-\!2u}}
\put(-0.09,-0.1){\spos{}{-\!(\!q\!-\!1\!)\l}}
\put(-0.05,-0.5){\spos{}{+\!\m}}\end{picture}}
\nc{\uuqll}{\begin{picture}(0,0)
\put(-0.1,0.3){\spos{}{-\!2u}}
\put(-0.09,-0.1){\spos{}{-\!(\!q\!-\!2\!)\l}}
\put(-0.05,-0.5){\spos{}{+\!\m}}\end{picture}}
\nc{\uuqql}{\begin{picture}(0,0)
\put(-0.1,0.3){\spos{}{-\!2u}}
\put(-0.06,-0.1){\spos{}{-\!(\!2q\!-\!1\!)\l}}
\put(-0.05,-0.5){\spos{}{+\!\m}}\end{picture}}
\nc{\uuqqll}{\begin{picture}(0,0)\put(-0.1,0.3){\spos{}{-\!2u}}
\put(-0.06,-0.1){\spos{}{-\!(\!2q\!-\!2\!)\l}}
\put(-0.05,-0.5){\spos{}{+\!\m}}\end{picture}}
\nc{\uuqqlll}{\begin{picture}(0,0)
\put(-0.1,0.3){\spos{}{-\!2u}}
\put(-0.06,-0.1){\spos{}{-\!(\!2q\!-\!3\!)\l}}
\put(-0.05,-0.5){\spos{}{+\!\m}}\end{picture}}
\nc{\um}{\begin{picture}(0,0)
\put(0.8,0){\spos{r}{-\!u\!+\!\m\;}}\end{picture}}
\nc{\uqm}{\begin{picture}(0,0)
\put(0.8,0){\spos{r}{-\!u\!-\!q\l\!+\!\m}}\end{picture}}
\nc{\uqlm}{\begin{picture}(0,0)
\put(0.8,0){\spos{r}{-\!u\!-\!(\!q\!-\!1\!)\l\!+\!\m}}
\end{picture}}
\nc{\uqllm}{\begin{picture}(0,0)
\put(0.8,0){\spos{r}{-\!u\!-\!(\!q\!-\!2\!)\l\!+\!\m}}
\end{picture}}
\nc{\luuq}{\begin{picture}(0,0)
\put(0,0){\spos{}{2u\!+\!q\l\!-\!\m}}\end{picture}}
\nc{\luuql}{\begin{picture}(0,0)
\put(0,0.3){\spos{}{2u}}
\put(-0.09,-0.1){\spos{}{+\!(\!q\!-\!1\!)\l}}
\put(-0.05,-0.5){\spos{}{-\!\m}}\end{picture}}
\nc{\luuqll}{\begin{picture}(0,0)
\put(0,0.3){\spos{}{2u}}
\put(-0.09,-0.1){\spos{}{+\!(\!q\!-\!2\!)\l}}
\put(-0.05,-0.5){\spos{}{-\!\m}}\end{picture}}
\nc{\luuqql}{\begin{picture}(0,0)
\put(0,0.3){\spos{}{2u}}
\put(-0.06,-0.1){\spos{}{+\!(\!2q\!-\!1\!)\l}}
\put(-0.05,-0.5){\spos{}{-\!\m}}\end{picture}}
\nc{\luuqqll}{\begin{picture}(0,0)
\put(0,0.3){\spos{}{2u}}
\put(-0.06,-0.1){\spos{}{+\!(\!2q\!-\!2\!)\l}}
\put(-0.05,-0.5){\spos{}{-\!\m}}\end{picture}}
\nc{\luuqqlll}{\begin{picture}(0,0)
\put(0,0.3){\spos{}{2u}}
\put(-0.06,-0.1){\spos{}{+\!(\!2q\!-\!3\!)\l}}
\put(-0.05,-0.5){\spos{}{-\!\m}}
\end{picture}}
\nc{\fu}{\begin{picture}(0,0)\put(0.75,0.5){\spos{}{u}}
\end{picture}}
\nc{\qll}{\begin{picture}(0,0)
\put(0.75,0.5){\spos{}{u\!+\!(\!q\!-\!2\!)\l}}
\end{picture}}
\nc{\ql}{\begin{picture}(0,0)
\put(0.75,0.5){\spos{}{u\!+\!(\!q\!-\!1\!)\l}}
\end{picture}}
\nc{\q}{\begin{picture}(0,0)\put(0.75,0.5){\spos{}{u\!+\!q\l}}
\end{picture}}
\nc{\pl}{\begin{picture}(0,0)
\put(0.75,0.5){\spos{}{u\!-\!(\!p\!-\!1\!)\l}}
\end{picture}}
\nc{\qpl}{\begin{picture}(0,0)\put(0.75,0.7){\spos{}{u\!+}}
\put(0.75,0.3){\spos{}{(\!q\!-\!p\!-\!1\!)\l}}\end{picture}}
\nc{\qp}{\begin{picture}(0,0)
\put(0.75,0.5){\spos{}{u\!+\!(\!q\!-\!p\!)\l}}\end{picture}}
\nc{\uqpl}{\begin{picture}(0,0)\put(0.75,0.7){\spos{}{u\!+}}
\put(0.75,0.3){\spos{}{(\!q\!-\!p\!+\!1\!)\l}}\end{picture}}
\nc{\fum}{\begin{picture}(0,0)
\put(0.75,0.5){\spos{}{-\!u\!+\!\m}}\end{picture}}
\nc{\umqll}{\begin{picture}(0,0)
\put(0.75,0.7){\spos{}{-\!u\!+\!\m\!-}}
\put(0.75,0.3){\spos{}{(\!q\!-\!2\!)\l}}\end{picture}}
\nc{\umql}{\begin{picture}(0,0)
\put(0.75,0.7){\spos{}{-\!u\!+\!\m\!-}}
\put(0.75,0.3){\spos{}{(\!q\!-\!1\!)\l}}\end{picture}}
\nc{\umq}{\begin{picture}(0,0)
\put(0.75,0.5){\spos{}{-\!u\!+\!\m\!-\!q\l}}\end{picture}}
\nc{\umpl}{\begin{picture}(0,0)
\put(0.75,0.7){\spos{}{-\!u\!+\!\m\!-}}
\put(0.75,0.3){\spos{}{(\!p\!-\!1\!)\l}}\end{picture}}
\nc{\umqplll}{\begin{picture}(0,0)
\put(0.75,0.7){\spos{}{-\!u\!+\!\m\!-}}
\put(0.75,0.3){\spos{}{(\!q\!+\!p\!-\!3\!)\l}}\end{picture}}
\nc{\umqpll}{\begin{picture}(0,0)
\put(0.75,0.7){\spos{}{-\!u\!+\!\m\!-}}
\put(0.75,0.3){\spos{}{(\!q\!+\!p\!-\!2\!)\l}}\end{picture}}
\nc{\umqpl}{\begin{picture}(0,0)
\put(0.75,0.7){\spos{}{-\!u\!+\!\m\!-}}
\put(0.75,0.3){\spos{}{(\!q\!+\!p\!-\!1\!)\l}}\end{picture}}
\rnc{\v}{\begin{picture}(0,0)\put(-0.23,0){\spos{r}{v}}
\end{picture}}
\nc{\vrl}{\begin{picture}(0,0)
\put(-0.8,0){\spos{l}{v\!+\!(\!r\!-\!1\!)\l}}
\end{picture}}
\nc{\vvrl}{\begin{picture}(0,0)\put(-0.1,0.3){\spos{}{-\!2v}}
\put(-0.09,-0.1){\spos{}{-\!(\!r\!-\!1\!)\l}}
\put(-0.05,-0.5){\spos{}{+\!\m}}
\end{picture}}
\nc{\uv}{\begin{picture}(0,0)\put(0,0){\spos{}{u\!-\!v}}
\end{picture}}
\nc{\uvq}{\begin{picture}(0,0)\put(0,0.25){\spos{}{u\!-\!v}}
\put(-0.05,-0.15){\spos{}{+\!(\!q\!-\!1\!)\l}}\end{picture}}
\nc{\uvr}{\begin{picture}(0,0)\put(0,0.25){\spos{}{u\!-\!v}}
\put(-0.05,-0.15){\spos{}{-\!(\!r\!-\!1\!)\l}}\end{picture}}
\nc{\uvqr}{\begin{picture}(0,0)\put(0,0.25){\spos{}{u\!-\!v}}
\put(-0.05,-0.15){\spos{}{+\!(\!q\!-\!r\!)\l}}\end{picture}}
\nc{\uvll}{\begin{picture}(0,0)\put(0,0.25){\spos{}{u\!-\!v}}
\put(-0.05,-0.15){\spos{}{+\!(\!q\!-\!2\!)\l}}\end{picture}}
\nc{\uvm}{\begin{picture}(0,0)
\put(-0.05,0){\spos{}{-\!u\!-\!v\!+\!\m}}
\end{picture}}
\nc{\uvqm}{\begin{picture}(0,0)
\put(-0.1,0.3){\spos{}{-\!u\!-\!v}}
\put(-0.06,-0.1){\spos{}{-\!(\!q\!-\!1\!)\l}}
\put(-0.05,-0.5){\spos{}{+\!\m}}
\end{picture}}
\nc{\uvrm}{\begin{picture}(0,0)
\put(-0.1,0.3){\spos{}{-\!u\!-\!v}}
\put(-0.06,-0.1){\spos{}{-\!(\!r\!-\!1\!)\l}}
\put(-0.05,-0.5){\spos{}{+\!\m}}
\end{picture}}
\nc{\uvqrm}{\begin{picture}(0,0)
\put(-0.1,0.4){\spos{}{-\!u\!-\!v}}
\put(-0.06,0){\spos{}{-\!(\!q\!+\!r\!-\!2\!)\l}}
\put(-0.05,-0.4){\spos{}{+\!\m}}
\end{picture}}
\nc{\uvllm}{\begin{picture}(0,0)
\put(-0.1,0.3){\spos{}{-\!u\!-\!v}}
\put(-0.06,-0.1){\spos{}{-\!(\!q\!-\!2\!)\l}}
\put(-0.05,-0.5){\spos{}{+\!\m}}
\end{picture}}
\nc{\face}[5]{\begin{picture}(1.6,1.6)
\multiput(0.3,0.3)(1,0){2}{\line(0,1){1}}
\multiput(0.3,0.3)(0,1){2}{\line(1,0){1}}
\put(0.26,0.26){\spos{tr}{#1}}\put(1.34,0.26){\spos{tl}{#2}}
\put(1.34,1.34){\spos{bl}{#3}}
\put(0.26,1.34){\spos{br}{#4}}
\put(0.8,0.8){\spos{}{#5}}\end{picture}}
\nc{\dface}[5]{\begin{picture}(1.6,1.6)
\multiput(0.3,0.8)(0.5,0.5){2}{\line(1,-1){0.5}}
\multiput(0.3,0.8)(0.5,-0.5){2}{\line(1,1){0.5}}
\put(0.24,0.8){\spos{r}{#1}}\put(0.8,0.24){\spos{t}{#2}}
\put(1.36,0.8){\spos{l}{#3}}
\put(0.8,1.36){\spos{b}{#4}}
\put(0.8,0.8){\spos{}{#5}}\end{picture}}
\nc{\lefttri}[4]{\begin{picture}(1.1,1.6)
\put(0.3,0.3){\line(0,1){1}}\put(0.3,0.3){\line(1,1){0.5}}
\put(0.3,1.3){\line(1,-1){0.5}}
\put(0.26,0.26){\spos{tr}{#1}}\put(0.86,0.8){\spos{l}{#2}}
\put(0.26,1.34){\spos{br}{#3}}
\put(0.38,0.8){\spos{l}{#4}}\end{picture}}
\nc{\righttri}[4]{\begin{picture}(1.1,1.6)
\put(0.8,0.3){\line(0,1){1}}\put(0.8,0.3){\line(-1,1){0.5}}
\put(0.8,1.3){\line(-1,-1){0.5}}
\put(0.84,0.26){\spos{tl}{#1}}\put(0.24,0.8){\spos{r}{#2}}
\put(0.84,1.34){\spos{bl}{#3}}
\put(0.72,0.8){\spos{r}{#4}}\end{picture}}
\nc{\bdoubface}[8]{\begin{picture}(1.6,2.6)
\multiput(0.3,0.3)(1,0){2}{\line(0,1){2}}
\multiput(0.3,0.3)(0,1){3}{\line(1,0){1}}\d{1.3}{1.3}
\put(0.26,0.26){\spos{tr}{#1}}\put(1.34,0.26){\spos{tl}{#2}}
\put(1.4,1.3){\spos{l}{#3}}\put(1.34,2.34){\spos{bl}{#4}}
\put(0.26,2.34){\spos{br}{#5}}\put(0.24,1.3){\spos{r}{#6}}
\put(0.8,0.8){\spos{}{#7}}\put(0.8,1.8){\spos{}{#8}}
\end{picture}}
\nc{\tdoubface}[8]{\begin{picture}(1.6,2.6)
\multiput(0.3,0.3)(1,0){2}{\line(0,1){2}}
\multiput(0.3,0.3)(0,1){3}{\line(1,0){1}}\d{0.3}{1.3}
\put(0.26,0.26){\spos{tr}{#1}}\put(1.34,0.26){\spos{tl}{#2}}
\put(1.36,1.3){\spos{l}{#3}}\put(1.34,2.34){\spos{bl}{#4}}
\put(0.26,2.34){\spos{br}{#5}}\put(0.2,1.3){\spos{r}{#6}}
\put(0.8,0.8){\spos{}{#7}}\put(0.8,1.8){\spos{}{#8}}
\end{picture}}
\nc{\sdoubface}[8]{\begin{picture}(2.6,1.6)
\multiput(0.3,0.3)(1,0){3}{\line(0,1){1}}
\multiput(0.3,0.3)(0,1){2}{\line(1,0){2}}\d{1.3}{1.3}
\put(0.26,0.26){\spos{tr}{#1}}\put(1.3,0.24){\spos{t}{#2}}
\put(2.34,0.26){\spos{tl}{#3}}\put(2.34,1.34){\spos{bl}{#4}}
\put(1.3,1.4){\spos{b}{#5}}\put(0.26,1.34){\spos{br}{#6}}
\put(0.8,0.8){\spos{}{#7}}\put(1.8,0.8){\spos{}{#8}}
\end{picture}}
\nc{\leftdoubtri}[7]{\begin{picture}(2.2,3.8)
\put(0.3,0.3){\line(0,1){3.2}}\put(1.9,1.9){\line(-1,-1){1.6}}
\put(1.9,1.9){\line(-1,1){1.6}}\put(0.3,1.9){\line(1,-1){0.8}}
\put(0.3,1.9){\line(1,1){0.8}}\d{1.1}{1.1}
\put(0.26,0.26){\spos{tr}{#1}}\put(0.24,1.9){\spos{r}{#1}}
\put(0.26,3.54){\spos{br}{#1}}
\put(1.14,2.74){\spos{bl}{#2}}\put(1.96,1.9){\spos{l}{#3}}
\put(1.14,1.06){\spos{tl}{#4}}
\put(0.36,1.1){\spos{l}{#5}}\put(0.36,2.7){\spos{l}{#6}}
\put(1.1,1.9){\spos{}{#7}}
\end{picture}}
\nc{\rightdoubtri}[7]{\begin{picture}(2.2,3.8)
\put(1.9,0.3){\line(0,1){3.2}}\put(0.3,1.9){\line(1,-1){1.6}}
\put(0.3,1.9){\line(1,1){1.6}}\put(1.9,1.9){\line(-1,-1){0.8}}
\put(1.9,1.9){\line(-1,1){0.8}}\d{1.1}{2.7}
\put(1.94,0.26){\spos{tl}{#1}}\put(1.96,1.9){\spos{l}{#1}}
\put(1.94,3.54){\spos{bl}{#1}}\put(1.06,2.74){\spos{br}{#2}}
\put(0.24,1.9){\spos{r}{#3}}\put(1.06,1.06){\spos{tr}{#4}}
\put(1.84,1.1){\spos{r}{#5}}\put(1.84,2.7){\spos{r}{#6}}
\put(1.1,1.9){\spos{}{#7}}\end{picture}}
\nc{\longlowblock}[2]{\begin{picture}(4,1)
\multiput(0,0)(0.6,0){2}{\line(0,1){1}}
\multiput(3.4,0)(0.6,0){2}{\line(0,1){1}}
\multiput(0,0)(0,0.5){3}{\line(1,0){4}}
\multiput(0.3,0.25)(3.4,0){2}{\spos{}{#1}}
\multiput(0.3,0.75)(3.4,0){2}{\spos{}{#2}}
\multiput(0,0.5)(0.6,0){2}{\spos{}{\bullet}}
\multiput(3.4,0.5)(0.6,0){2}{\spos{}{\bullet}}
\end{picture}}
\nc{\shortlowblock}[2]{\begin{picture}(2,1)
\multiput(0,0)(0.6,0){2}{\line(0,1){1}}
\multiput(1.4,0)(0.6,0){2}{\line(0,1){1}}
\multiput(0,0)(0,0.5){3}{\line(1,0){2}}
\multiput(0.3,0.25)(1.4,0){2}{\spos{}{#1}}
\multiput(0.3,0.75)(1.4,0){2}{\spos{}{#2}}
\multiput(0,0.5)(0.6,0){2}{\spos{}{\bullet}}
\multiput(1.4,0.5)(0.6,0){2}{\spos{}{\bullet}}
\end{picture}}
\nc{\leftlowend}[1]{\begin{picture}(0.5,1)
\put(0,0){\line(0,1){1}}\put(0,0){\line(1,1){0.5}}
\put(0,1){\line(1,-1){0.5}}
\put(0.04,0.5){\spos{l}{#1}}
\multiput(0,0)(0,1){2}{\makebox(0.5,0){\dotfill}}
\end{picture}}
\nc{\rightlowend}[1]{\begin{picture}(0.5,1)
\put(0,0.5){\line(1,-1){0.5}}\put(0,0.5){\line(1,1){0.5}}
\put(0.5,0){\line(0,1){1}}
\put(0.46,0.5){\spos{r}{#1}}
\multiput(0,0)(0,1){2}{\makebox(0.5,0){\dotfill}}\end{picture}}
\nc{\leftbu}[2]{\begin{picture}(1.5,1)
\put(0,0){\line(0,1){1}}\put(0,0){\line(1,1){1}}
\put(0,1){\line(1,-1){1}}\put(1,0){\line(1,1){0.5}}
\put(1,1){\line(1,-1){0.5}}
\put(0.04,0.5){\spos{l}{#1}}\put(1,0.5){\spos{}{#2}}
\multiput(0,0)(0,1){2}{\makebox(1.5,0){\dotfill}}
\put(0.5,0.5){\spos{}{\bullet}}\end{picture}}
\nc{\rightbu}[2]{\begin{picture}(1.5,1)
\put(0,0.5){\line(1,-1){0.5}}\put(0,0.5){\line(1,1){0.5}}
\put(0.5,1){\line(1,-1){1}}\put(0.5,0){\line(1,1){1}}
\put(1.5,0){\line(0,1){1}}
\put(0.5,0.5){\spos{}{#1}}\put(1.46,0.5){\spos{r}{#2}}
\multiput(0,0)(0,1){2}{\makebox(1.5,0){\dotfill}}
\put(1,0.5){\spos{}{\bullet}}\end{picture}}
\nc{\longblock}[4]{\begin{picture}(4,2)
\multiput(0,0)(0.6,0){2}{\line(0,1){2}}
\multiput(3.4,0)(0.6,0){2}{\line(0,1){2}}
\multiput(0,0)(0,0.5){5}{\line(1,0){4}}
\multiput(0.3,0.25)(3.4,0){2}{\spos{}{#1}}
\multiput(0.3,0.75)(3.4,0){2}{\spos{}{#2}}
\multiput(0.3,1.25)(3.4,0){2}{\spos{}{#3}}
\multiput(0.3,1.75)(3.4,0){2}{\spos{}{#4}}
\multiput(0,0.5)(0,0.5){3}{\spos{}{\bullet}}
\multiput(0.6,0.5)(0,0.5){3}{\spos{}{\bullet}}
\multiput(3.4,0.5)(0,0.5){3}{\spos{}{\bullet}}
\multiput(4,0.5)(0,0.5){3}{\spos{}{\bullet}}
\end{picture}}
\nc{\shortblock}[4]{\begin{picture}(2,2)
\multiput(0,0)(0.6,0){2}{\line(0,1){2}}
\multiput(1.4,0)(0.6,0){2}{\line(0,1){2}}
\multiput(0,0)(0,0.5){5}{\line(1,0){2}}
\multiput(0.3,0.25)(1.4,0){2}{\spos{}{#1}}
\multiput(0.3,0.75)(1.4,0){2}{\spos{}{#2}}
\multiput(0.3,1.25)(1.4,0){2}{\spos{}{#3}}
\multiput(0.3,1.75)(1.4,0){2}{\spos{}{#4}}
\multiput(0,0.5)(0,0.5){3}{\spos{}{\bullet}}
\multiput(0.6,0.5)(0,0.5){3}{\spos{}{\bullet}}
\multiput(1.4,0.5)(0,0.5){3}{\spos{}{\bullet}}
\multiput(2,0.5)(0,0.5){3}{\spos{}{\bullet}}
\end{picture}}
\nc{\inv}[2]{\begin{picture}(2,1)
\put(0.5,0){\line(1,1){1}}\put(0.5,1){\line(1,-1){1}}
\multiput(0,0.5)(1.5,0.5){2}{\line(1,-1){0.5}}
\multiput(0,0.5)(1.5,-0.5){2}{\line(1,1){0.5}}
\put(0.5,0.5){\spos{}{#1}}\put(1.5,0.5){\spos{}{#2}}
\put(1,0.5){\spos{}{\bullet}}
\multiput(0,0)(0,1){2}{\makebox(2,0){\dotfill}}
\end{picture}}
\nc{\lowinv}[2]{\begin{picture}(2,2)
\put(0,0){\inv{#1}{#2}}
\multiput(0,1.5)(0,0.5){2}{\makebox(2,0){\dotfill}}
\multiput(0.5,1)(1,0){2}{\spos{}{\bullet}}
\end{picture}}
\nc{\midinv}[2]{\begin{picture}(2,2)
\put(0,0.5){\inv{#1}{#2}}
\multiput(0,0)(0,2){2}{\makebox(2,0){\dotfill}}
\multiput(0.5,0.5)(1,0){2}{\spos{}{\bullet}}
\multiput(0.5,1.5)(1,0){2}{\spos{}{\bullet}}
\end{picture}}
\nc{\highinv}[2]{\begin{picture}(2,2)
\put(0,1){\inv{#1}{#2}}
\multiput(0,0)(0,0.5){2}{\makebox(2,0){\dotfill}}
\multiput(0.5,1)(1,0){2}{\spos{}{\bullet}}
\end{picture}}
\nc{\leftend}[2]{\begin{picture}(0.5,2)
\put(0,0){\line(0,1){2}}\multiput(0,0)(0,1){2}{\line(1,1){0.5}}
\multiput(0,1)(0,1){2}{\line(1,-1){0.5}}
\put(0.04,0.5){\spos{l}{#1}}\put(0.04,1.5){\spos{l}{#2}}
\multiput(0,0)(0,1){3}{\makebox(0.5,0){\dotfill}}
\put(0,1){\spos{}{\bullet}}
\end{picture}}
\nc{\rightend}[2]{\begin{picture}(0.5,2)
\put(0.5,0){\line(0,1){2}}\multiput(0.5,0)(0,1){2}{\line(-1,1){0.5}}
\multiput(0.5,1)(0,1){2}{\line(-1,-1){0.5}}
\put(0.46,0.5){\spos{r}{#1}}\put(0.46,1.5){\spos{r}{#2}}
\multiput(0,0)(0,1){3}{\makebox(0.5,0){\dotfill}}
\put(0.5,1){\spos{}{\bullet}}
\end{picture}}
\nc{\lefttriangle}[3]{\begin{picture}(1,2)
\put(0,0){\line(0,1){2}}\put(0,1){\line(1,1){0.5}}
\put(0,1){\line(1,-1){0.5}}\put(1,1){\line(-1,1){1}}
\put(1,1){\line(-1,-1){1}}
\put(0.04,0.5){\spos{l}{#1}}\put(0.04,1.5){\spos{l}{#2}}
\put(0.5,1){\spos{}{#3}}
\multiput(0,0)(0,2){2}{\makebox(1,0){\dotfill}}
\multiput(0.5,0.5)(0,1){2}{\makebox(0.5,0){\dotfill}}
\put(0,1){\spos{}{\bullet}}
\multiput(0.5,0.5)(0,1){2}{\spos{}{\bullet}}
\end{picture}}
\nc{\righttriangle}[3]{\begin{picture}(1,2)
\put(1,0){\line(0,1){2}}\put(1,1){\line(-1,1){0.5}}
\put(1,1){\line(-1,-1){0.5}}\put(0,1){\line(1,1){1}}
\put(0,1){\line(1,-1){1}}
\put(0.96,0.5){\spos{r}{#1}}\put(0.96,1.5){\spos{r}{#2}}
\put(0.5,1){\spos{}{#3}}
\multiput(0,0)(0,2){2}{\makebox(1,0){\dotfill}}
\multiput(0,0.5)(0,1){2}{\makebox(0.5,0){\dotfill}}
\put(1,1){\spos{}{\bullet}}
\multiput(0.5,0.5)(0,1){2}{\spos{}{\bullet}}
\end{picture}}
\nc{\botleftref}[4]{\begin{picture}(1.5,2)
\put(0,0){\line(0,1){2}}\put(0,1){\line(1,1){0.5}}
\put(0,1){\line(1,-1){1}}\put(0,0){\line(1,1){1}}
\put(0,2){\line(1,-1){1.5}}\put(1.5,0.5){\line(-1,-1){0.5}}
\put(0.04,0.5){\spos{l}{#1}}\put(0.04,1.5){\spos{l}{#2}}
\put(1,0.5){\spos{}{#3}}\put(0.5,1){\spos{}{#4}}
\multiput(0,0)(0,2){2}{\makebox(1.5,0){\dotfill}}
\put(1,1){\makebox(0.5,0){\dotfill}}
\put(0.5,1.5){\makebox(1,0){\dotfill}}
\multiput(0,1)(1,0){2}{\spos{}{\bullet}}
\multiput(0.5,0.5)(0,1){2}{\spos{}{\bullet}}
\end{picture}}
\nc{\topleftref}[4]{\begin{picture}(1.5,2)
\put(0,0){\line(0,1){2}}\put(0,1){\line(1,1){1}}
\put(0,1){\line(1,-1){0.5}}\put(0,0){\line(1,1){1.5}}
\put(0,2){\line(1,-1){1}}\put(1.5,1.5){\line(-1,1){0.5}}
\put(0.04,0.5){\spos{l}{#1}}\put(0.04,1.5){\spos{l}{#2}}
\put(0.5,1){\spos{}{#3}}\put(1,1.5){\spos{}{#4}}
\multiput(0,0)(0,2){2}{\makebox(1.5,0){\dotfill}}
\put(1,1){\makebox(0.5,0){\dotfill}}
\put(0.5,0.5){\makebox(1,0){\dotfill}}
\multiput(0,1)(1,0){2}{\spos{}{\bullet}}
\multiput(0.5,0.5)(0,1){2}{\spos{}{\bullet}}
\end{picture}}
\nc{\botrightref}[4]{\begin{picture}(1.5,2)
\put(1.5,0){\line(0,1){2}}\put(1.5,1){\line(-1,1){0.5}}
\put(1.5,1){\line(-1,-1){1}}\put(1.5,0){\line(-1,1){1}}
\put(1.5,2){\line(-1,-1){1.5}}\put(0,0.5){\line(1,-1){0.5}}
\put(1.46,0.5){\spos{r}{#1}}\put(1.46,1.5){\spos{r}{#2}}
\put(0.5,0.5){\spos{}{#3}}\put(1,1){\spos{}{#4}}
\multiput(0,0)(0,2){2}{\makebox(1.5,0){\dotfill}}
\put(0,1){\makebox(0.5,0){\dotfill}}
\put(0,1.5){\makebox(1,0){\dotfill}}
\multiput(0.5,1)(1,0){2}{\spos{}{\bullet}}
\multiput(1,0.5)(0,1){2}{\spos{}{\bullet}}
\end{picture}}
\nc{\toprightref}[4]{\begin{picture}(1.5,2)
\put(1.5,0){\line(0,1){2}}\put(1.5,1){\line(-1,1){1}}
\put(1.5,1){\line(-1,-1){0.5}}\put(1.5,0){\line(-1,1){1.5}}
\put(1.5,2){\line(-1,-1){1}}\put(0,1.5){\line(1,1){0.5}}
\put(1.46,0.5){\spos{r}{#1}}\put(1.46,1.5){\spos{r}{#2}}
\put(1,1){\spos{}{#3}}\put(0.5,1.5){\spos{}{#4}}
\multiput(0,0)(0,2){2}{\makebox(1.5,0){\dotfill}}
\put(0,1){\makebox(0.5,0){\dotfill}}
\put(0,0.5){\makebox(1,0){\dotfill}}
\multiput(0.5,1)(1,0){2}{\spos{}{\bullet}}
\multiput(1,0.5)(0,1){2}{\spos{}{\bullet}}
\end{picture}}

\begin{document}

 \title{Finitized Conformal Spectra of the Ising Model\\
      on the Klein Bottle and M\"obius Strip}

    \author{C. H. Otto Chui\thanks{
    e-mail: {\tt C.Chui@ms.unimelb.edu.au}} \ and Paul
    A.~Pearce\thanks{ e-mail: {\tt P.Pearce@ms.unimelb.edu.au} }
    \\[2pt]
    \it Department of Mathematics and Statistics\\[-8pt]
    \it University of Melbourne\\[-8pt]
    \it Parkville, Victoria 3010, Australia}

    \date{May, 2001} 
    \maketitle
    %

    \begin{abstract}
    We study the conformal spectra of the critical square lattice Ising
    model on the Klein bottle and M\"obius strip using Yang-Baxter
    techniques and the solution of functional equations. In particular, we
    obtain expressions for the {\it finitized} conformal  partition 
    functions in terms
    of {\it finitized} Virasoro characters.  This demonstrates that Yang-Baxter
    techniques  and functional equations can be used to study the conformal spectra
    of more general exactly solvable lattice models in these topologies. 
    The results rely on certain
    properties of the eigenvalues which are confirmed numerically.
    \end{abstract}


\nsection{Introduction}

There has been much recent progress on understanding general conformal
boundary conditions for rational theories on the cylinder~\cite{BPPZ} and
torus~\cite{PetZuber} and their relation~\cite{BP2} to
integrable boundary conditions on the lattice. It is therefore of 
some interest to
extend this understanding to other topologies such as the Klein 
bottle and M\"obius
strip. However, while much is known~\cite{PSS,SS,HSS} about conformal partition
functions in these topologies from the viewpoint of string theory, very
little is known  about integrable lattice boundary conditions on the
Klein bottle and M\"obius strip. Indeed, the lattice Yang-Baxter 
techniques which
are well developed for the torus and cylinder have not yet been exploited in
other topologies.

In this paper, we use Yang-Baxter techniques to study the conformal spectra of
the critical square lattice Ising model on the Klein bottle and M\"obius strip
shown in Figure~1. Although the Ising model has been
studied previously in these topologies by free-fermion~\cite{Yamaguchi} and
Pfaffian~\cite{LuWu1,LuWu2,LuWu3} techniques, we point out that these 
techniques do not
generalize to other exactly solvable lattice models. In contrast, the methods
based on Yang-Baxter techniques and the solution of functional 
equations developed
in this paper will generalize to other exactly solvable lattice such as the
$A$-$D$-$E$ lattice models \cite{Pasquier}.

\begin{figure}[t]
\setlength{\unitlength}{7.5mm}
\begin{picture}(4,6)
\put(1.5,0){
\begin{picture}(5,5)
\multiput(1,1)(0,1){3}{\line(1,0){5}}
\multiput(1,5)(0,1){3}{\line(1,0){5}}
\multiput(1,1)(1,0){2}{\line(0,1){6}}
\multiput(5,1)(1,0){2}{\line(0,1){6}}
\multiput(1.15,1.15)(0,1){2}{\pp{}{\searrow}}
\multiput(5.15,1.15)(0,1){2}{\pp{}{\searrow}}
\multiput(1.15,5.15)(0,1){2}{\pp{}{\searrow}}
\multiput(5.15,5.15)(0,1){2}{\pp{}{\searrow}}
\multiput(1.5,1.5)(4,0){2}{\pp{c}{u}}
\multiput(1.5,2.5)(4,0){2}{\pp{c}{\lambda-u}}
\multiput(1.5,5.5)(4,0){2}{\pp{c}{u}}
\multiput(1.5,6.5)(4,0){2}{\pp{c}{\lambda-u}}
\multiput(3,0.85)(0,6){1}{$\gg$}
\multiput(3,6.85)(0,0){1}{$\ll$}
\multiput(0.8,4)(5,0){2}{$\wedge$}
\end{picture}}
%
%
\put(10.5,0){
\begin{picture}(5,5)
\multiput(1,1)(0,1){3}{\line(1,0){5}}
\multiput(1,5)(0,1){3}{\line(1,0){5}}
\multiput(1,1)(5,0){2}{\thicklines{\line(0,1){6}} }
\multiput(2,1)(3,0){2}{\line(0,1){6}}
\multiput(1.5,1.5)(4,0){2}{\pp{c}{u}}
\multiput(1.5,2.5)(4,0){2}{\pp{c}{\lambda-u}}
\multiput(1.5,5.5)(4,0){2}{\pp{c}{u}}
\multiput(1.5,6.5)(4,0){2}{\pp{c}{\lambda-u}}
\multiput(1.15,1.15)(0,1){2}{\pp{}{\searrow}}
\multiput(5.15,1.15)(0,1){2}{\pp{}{\searrow}}
\multiput(1.15,5.15)(0,1){2}{\pp{}{\searrow}}
\multiput(5.15,5.15)(0,1){2}{\pp{}{\searrow}}
\multiput(3,0.85)(0,6){1}{$\gg$}
\multiput(3,6.85)(0,0){1}{$\ll$}
\end{picture}}
\end{picture}
\caption{Klein bottle (left) \& M\"obius strip (right). The boundaries are joined according to the orientations indicated by the arrows. The
arrows at the bottom left corners on each face weight indicate its orientation.}
\label{KleinFig}
\end{figure}
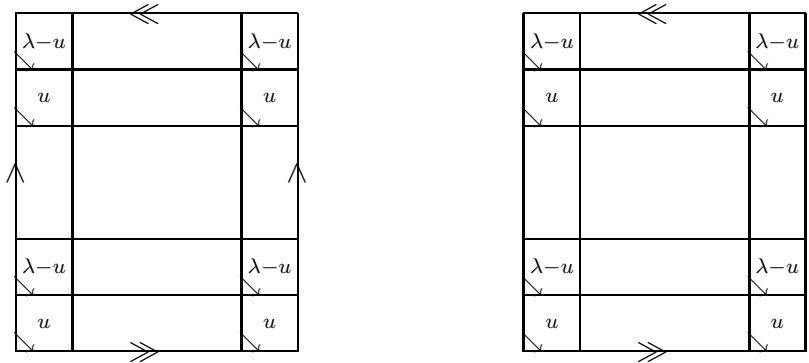
%
%

The layout of the paper is as follows. In Section~2 we review the Ising
model on the torus. We define the $A_3$ representation of the Ising
model and its single row transfer matrix $\T(u)$. We also summarize the
solution~\cite{Baxter,OPW} for the {\it finitized} conformal partition function
using functional equations. In Section~3, we consider the Ising model 
on the Klein
bottle.  We define the double row transfer matrices
$\D(u)=\T(u)\T(\lambda-u)$ and the flip
operator $\F$ and discuss the simultaneous eigenvectors of $\F$ and $\D(u)$
and their relation to the eigenvectors of $\T(u)$.
We calculate the
conformal spectra of the Ising model on the Klein bottle in terms of
{\it finitized} Virasoro characters by solving the relevant functional equation
in the form of an inversion identity. In Section~4 we consider the Ising
model on the M\"obius strip. We define the double row transfer matrices
$\D(u)$ with integrable boundary conditions corresponding to $+$ and free
boundaries. The relevant inversion identities for the conformal spectra of
the Ising model on the M\"obius strip are the same
as for the cylinder with $+$ or free boundary conditions on the left and right.
For each eigenvalue, we identify the eigenvalue $F=\pm 1$ under the 
flip operator
$\F$ according to the classifying pattern of 1- and 2-strings in the complex
$u$-plane. This identification is not obtained analytically but is confirmed
by extensive numerics determining the parity $F=\pm 1$ of the associated
eigenvectors. Finally, we end with a brief discussion.

\nsection{Ising Model on the Torus}\label{secfun}
In this section we briefly review the critical Ising model with periodic
boundary  conditions on the torus and its finitized partition
function~\cite{FF,Baxter,OPW}.

\subsection{Ising model}\label{Isingusual}

The Hamiltonian of the Ising model is given by
\begin{equation}
H(\{\sigma\})\:=\:-\sum_{<i,j>}J_{ij}\sigma_i\sigma_j -h \sum_i\sigma_i
\end{equation} 
where $\sigma_i=\pm1$ is the spin at site $i$, $J_{ij}\ge0$ the interaction strength between site $i$ and $j$, and $h$ the external field. The sum is over all pairs of sites $<\!i,j\!>$.

In this article, we consider only the nearest-neighbour critical Ising 
model on square lattice with zero external field. The partition function is given by
\begin{equation}
Z(K,L)\:=\:\sum_{\{\sigma\}}\exp(K\!\!\sum_{<i,j>_h}\sigma_i\sigma_j+
L\!\!\sum_{<i,j>_v}\sigma_i\sigma_j ) \label{usualpf}
\end{equation}   
where $K,L$ are the couplings in horizontal and vertical direction respectively. The model is critical on the line \,$\sinh(2K)\sinh(2L)\!=\!1$. The partition function is a sum over all possible configurations of spins on the lattice with/without constraints due to the geometry of the lattice and the boundary conditions.

However, since the row-to-row transfer matrices in this representation do not commute, Baxter's commuting transfer matrices technique cannot be applied directly. 
Hence, we work in the $A_3$ representaion defined in the next sub-section
for the following reasons:
\begin{enumerate}
\item  
The critical Ising model on a square lattice can be conveniently
formulated as the critical $A_3$ model.
\item It can be easily generalised to other $A\!-\!D\!-\!E$ models.
\end{enumerate}

\subsection{$A_3$ representation and its transfer matrices}\label{secA3}

The $A_3$ model is one of the 
Andrews-Baxter-Forrester models~\cite{ABF} $A_L$ with $L=3$
introduced in 1984.
In 1987, Pasquier constructed the critical $A\!-\!D\!-\!E$ models
\cite{Pasquier}
which include the critical $A_L$ models.
In the $A_3$ model, the spins
$a,b,c,d,\ldots$ assigned to the sites of the lattice take heights from the
set $\{1,2,3\}$ and satisfy the adjacency condition that heights on
adjacent sites must differ by $\pm1$. 

Consider the Ising model on a square lattice of $N$ columns and $M$ rows where
$N$=$2L$ and $M$=$2L'$ are both even.

The Boltzmann face weights of the critical $A_3$ model are
\setlength{\unitlength}{13mm}
\begin{equation}
\!\!\!\!\!\!\!\!\!{\W{d}{c}{u}{a}{b}}
= 
\begin{picture}(1.14,1)
\put(-0.2,-0.7){
\begin{picture}(1.14,1)
\put(0.57,0.8){\setlength{\unitlength}{0.71\unitlength}
\makebox(0,0){\face{a}{b}{c}{d}{u}}}
\put(0.4,0.53){\pp{}{\searrow}}
\end{picture} }
\end{picture}
=\frac{\sin(\lambda-u)}{\sin \lambda}\;\delta_{a,c}+\frac{\sin
   u}{\sin \lambda}\left(\frac{S_aS_c}{S_bS_d}\right)^{1/2}
\delta_{b,d} \label{BoltzA3}
\end{equation}
%
where $\lambda=\frac{\pi}{4}$ is the crossing parameter and
$S_a=\sin(a\lambda)$ are the crossing factors.
It is understood that the face weights vanish if the adjacency constraint
$a-b=\pm 1$ is not satisfied along any edge. The spectral parameter $u$ is
usually taken in the range $0<u<\lambda$, so that the face weights are all
positive but can be extended into the complex $u$-plane by analytic
continuation.

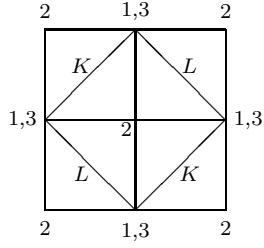
\begin{figure}[t]
\setlength{\unitlength}{12mm}
\begin{picture}(3,3)
\put(4,0){
\begin{picture}(3,3)
\multiput(1,1)(1,0){3}{\line(0,1){2}}
\multiput(1,1)(0,1){3}{\line(1,0){2}}
\multiput(2,1)(1,1){2}{\line(-1,1){1}}
\multiput(2,1)(-1,1){2}{\line(1,1){1}}
\multiput(1,0.8)(2,0){2}{\pp{}{2}}
\multiput(1,3.2)(2,0){2}{\pp{}{2}}
\put(1.9,1.9){\pp{}{2}}
\multiput(0.75,2)(2.5,0){2}{\pp{}{1,3}}
\multiput(2,0.75)(0,2.45){2}{\pp{}{1,3}}
\multiput(1.4,1.4)(1.2,1.2){2}{\pp{}{L}}
\multiput(2.6,1.4)(-1.2,1.2){2}{\pp{}{K}}
\end{picture} }
\end{picture}
\mbox{}\vspace{-1cm}\mbox{}
\caption{Ising model couplings on $A_3$ lattice.}
\label{A3Isingequiv}
\end{figure}

The $A_3$ model corresponds to the two dimensional Ising model.
In $A_3$ model, the square lattice has two sublattices which can
be even or odd. 
The heights $a$ are even on the even sublattice 
and odd on the odd sublattice. 
On the even sublattice, the heights are fixed to the value 2. 
On the odd sublattice we identify the state
$a=1$ with the usual $+$ Ising state and $a=3$ with the usual $-$ 
Ising state. 
By the adjacency condition, if one of the sublattice is odd, then the other
sublattice must be even and vice versa.
The Ising model couplings act along the diagonals of the faces as depicted in
fig.$\,$\ref{A3Isingequiv}.
It can be shown that the spectral parameter $u$ is related to the couplings by 
\begin{equation}
\begin{array}{l}
e^{2K}=\W 2 1 u 1 2 / \W 2 3 u 1 2=\frac{\sin(2\lambda-u)}{\sin u} \\
e^{2L}=\W 1 2 u 2 1 / \W 3 2 u 2 1=\frac{\sin(\lambda+u)}{\sin(\lambda-u)}
\end{array}
\end{equation}
It is obvious that  the interactions are isotropic at
$u=\frac{\lambda}{2}$.
From this, we see that the $A_3$ model corresponds to the Ising model.
Note that in the $A_3$ representation, it contains two mutually independent copies of Ising models.
From now on, without stating explicitly, we use the term ``Ising model'' meaning
``the Ising model in $A_3$ representation''.

The face weights \eqref{BoltzA3} have
the following local properties.
They are symmetric under reflection about the
diagonals
\begin{equation}
\W d c u a b = \W bcuad=\W daucb \label{diag}
\end{equation}
and satisfy the crossing symmetries
\begin{equation}
\W d c u a b=\left({{S_a S_c}\over {S_b S_d}}\right)^{1/2}
\!\!\W c b{\!\lambda-u} d a \label{crossym}
= \left({{S_a S_c}\over {S_b S_d}}\right)^{1/2}
\!\!\W a d{\!\lambda-u} b c
\end{equation}
 By \eqref{diag} and \eqref{crossym}, we also have the following
symmetries under horizontal and vertical reflections
\begin{equation}
\W d c u a b =\left({{S_a S_c}\over {S_b S_d}}\right)^{1/2}
\!\!\W a b {\!\lambda-u} d c \label{reflsym}
= \left({{S_a S_c}\over {S_b S_d}}\right)^{1/2}
\!\!\W c d {\!\lambda-u} b a \label{reflectweight}
\end{equation}

The single row transfer matrix $\T(u)$ between the rows of heights
$\a=\{a_1,\cdots,a_N\}$ and $\b=\{b_1,\cdots,b_N\}$ has entries
\begin{equation}
\T(u)_{\smb a,\smb b}=\prod_{j=1}^{N} \W {b_j} {b_{j+1}} u {a_j}
{a_{j+1}}
\end{equation}
with the periodicity $a_{N+1}=a_1$, $b_{N+1}=b_1$.

Since the Boltzmann weights satisfy the Yang-Baxter equation
and inversion relation,
the transfer matrices $\T(u)$ form a commuting family~\cite{Baxter,BPO} with
$\T(u)\T(v)=\T(v)\T(u)$. 
Using \eqref{reflsym} and the periodic boundary
condition along the row we find
%
%
\begin{equation}
\T(\lambda-u)_{\smb a,\smb b}=\T(u)_{\smb b,\smb a}=[\T^T(u)]_{\smb a,\smb b}
\end{equation}
So $\T(\lambda-u)=\T^T(u)$ and $\T(u)$ both belong to a commuting family of
normal matrices which can be simultaneously diagonalized by a unitary 
matrix. The
corresponding eigenvectors $\{\z_n\}$ are in general complex and independent of
$u$.

The adjacency matrix for the Ising model in the $A_3$ representation is
\begin{equation}
A=\mbox{\small $\pmatrix{0&1&0\cr 1&0&1\cr 0&1&0}$}
\label{diagA}
\end{equation}
i.\ e.\ $A_{ij}=1$ if heights $i$ \& $j$ are admissible on adjacent sites and zero otherwise.
It follows that the dimension of the transfer matrix is
\begin{equation}
\mbox{dim}\,\T(u)=\mbox{Tr}\,A^{N}=2(2^{N/2})=2^{L+1}
\end{equation}
Note that the dimension is twice larger than the one in usual representation as defined in section \ref{Isingusual} since we are working in $A_3$ representation which contains two copies of Ising models.
In fact, we can order all the row configurations into two blocks. In the first
block the first height in each row is even. In the second block, we cyclically
translate the row configurations of the first block by one unit to the right so
that the first height in each row is now odd. Using the fact that 
$\T(u)$ commutes
with the shift operator $\C=\T(0)$ which is defined as
\begin{equation}
\C_{\smb a,\smb b}=\T(0)_{\smb a,\smb b}=\prod_{j=1}^N \delta_{a_j,b_{j+1}}
\end{equation}
 we obtain, in this basis, the block matrices
\begin{equation}
\C=\pmatrix{\mathbf 0&\I\cr \tilde{\C}&\mathbf 0},\qquad
\T(u)=\pmatrix{
{\mathbf 0}& \V(u)\cr
\tilde{\C}\V(u)& {\mathbf 0}
}
\label{CTbasis}
\end{equation}
where the orthogonal shift matrix $\tilde{\C}$ commutes with $\V(u)$
and each of the two blocks are of size $2^L$.
It follows that the eigenvalues of $\T(u)$ occur in pairs $\pm T(u)$.
As an aside, $\V(u)$ and $\tilde{\C}\V(u)$  in \eqref{CTbasis},
up to normalization, are equal to $\V(u)$
and $\mathbf{W}(u)$ as defined in chapter~7 of \cite{Baxter} respectively.

\subsection{Finitized Ising partition function on the torus}

The critical properties of the Ising model are described by the unitary minimal
conformal field theory with central charge $c=1/2$ and conformal
weights $\Delta=0,1/2$ and $1/16$. This conformal data is related to
lattice quantities through finite-size corrections~\cite{BCN,Affleck} to the
eigenvalues of the transfer matrices.

Consider the finite-size partition function $Z_{NM}$ of the
critical Ising model on a periodic lattice of $N$ columns and
$M$ rows. The asymptotic behaviour of  $Z_{NM}$  in the
limit of large $N$ and $M$ with the aspect ratio $M/N$ fixed is given by
\begin{equation}
Z_{NM}^\sTorus(u)=\mbox{Tr} [\T(u)^M]\sim
\exp\left[-NM\fb(u)\right]\, Z^{\sTorus}(q) \label{ZNM}
\end{equation}
where $\T(u)$ is the periodic row transfer matrix, $\fb(u)$ is the bulk free
energy and $Z^\sTorus(q)$ is the universal conformal
partition function with  modular parameter
\begin{equation}
q=\exp\Bigl(-2\pi i e^{-4iu}\, {M\over N}\Bigr). \label{modtor}
\end{equation}
Since the lattice is periodic, there is no boundary free energy.

The finite-size corrections of the eigenvalues $T(u)$ of the Ising 
model on a torus
are given by~\cite{BCN,Affleck}
\begin{equation}
\log T(u) = -N\fb+\frac{2\pi}{N} \left[ \Bigl(\frac{c}{12}-
\Delta-\overline{\Delta}\Bigr)\sin 4u -ik e^{-4iu} + i\bar{k}
e^{4iu} \right] +\mbox{o}\Bigl(\frac{1}{N}\Bigr) \label{pcorr}
\end{equation}
where $\Delta,\overline{\Delta}$ are conformal weights and $k$ and $\bar{k}$
are arbitrary non-negative integers yielding towers of eigenvalues above each primary
level $(\Delta,\overline{\Delta})$ with $k=\overline{k}=0$.
The modular invariant partition function is expressed in terms of
Virasoro characters as
\begin{eqnarray}
Z^\sTorus(q)&=& \sum_{(\Delta,\overline{\Delta})}
{\cal N}(\Delta,\overline{\Delta})\; q^{-c/12+\Delta+\overline{\Delta}}
\sum_k q^{k}
\sum_{\bar{k}} \overline{q}\,^{\bar{k}}\\
&=&\chi_0(q)\chi_0(\bar q) +
\chi_{1/16}(q)\chi_{1/16}(\bar q)
  + \chi_{1/2}(q)\chi_{1/2}(\bar q)\nonumber \label{mipf}
\end{eqnarray}
where $\overline q$ is the complex conjugate of $q$ and the operator content is
given by
\begin{equation}
{\cal N}(\Delta,\overline{\Delta})=
\cases{1, &$(\Delta,\overline{\Delta})=(0,0),(1/16,1/16),(1/2,1/2)$\cr
0,& otherwise}
\end{equation}

The $c=1/2$ Virasoro characters are given variously by
\begin{eqnarray}
q^{1/48}\chi_0(q)&=&\frac{1}{2} \left\{ \prod_{k=1}^{\infty}
\left(1+q^{k-1/2}\right)+ \prod_{k=1}^{\infty}
\left(1-q^{k-1/2}\right) \right\}\\
&=& \hspace{-4mm} \sum_{ \renewcommand{\arraystretch}{0.6}
\begin{array}{c} \sc m\geq 0\\ \sc m \smbox{ even} \end{array}}
\hspace{-1mm} \frac{q^{\case{1}{2} m^2}}{(q)_m} \; = \;
\frac{1}{(q)_{\infty}} \sum_{j=-\infty}^{\infty} \left\{
q^{j(12j+1)}-q^{(3j+1)(4j+1)}\right\} \label{chi0}\nonumber \\
q^{-23/48} \chi_{1/2}(q)&=&\frac{q^{-1/2}}{2} \left\{
\prod_{k=1}^{\infty} \left(1+q^{k-1/2}\right)-
\prod_{k=1}^{\infty} \left(1-q^{k-1/2}\right) \right\}\\
&=& \hspace{-4mm} \sum_{ \renewcommand{\arraystretch}{0.6}
\begin{array}{c} \sc m\geq 0\\ \sc m \smbox{ odd} \end{array}}
\hspace{-1mm} \frac{q^{\case{1}{2} (m^2-1)}}{(q)_m} \; = \;
\frac{1}{(q)_{\infty}} \sum_{j=-\infty}^{\infty}\!\! \left\{
q^{j(12j+5)}-q^{(3j+2)(4j+1)}\right\} \label{chi12}\nonumber \\
q^{-1/24} \chi_{1/16}(q) &=&
\prod_{k=1}^{\infty} \left(1+q^k\right) \label{chi16} \\
&=& \sum_{m\geq 0} \frac{q^{\case{1}{2} m(m+1)}}{(q)_m} \; =
\; \frac{1}{(q)_{\infty}} \sum_{j=-\infty}^{\infty} \left\{
q^{j(12j-2)}-q^{(3j+1)(4j+2)}\right\}\nonumber
\end{eqnarray}
where $(q)_m = \prod_{k=1}^m (1-q^k)$ for $m>0$ and $(q)_0=1$.
The three different forms of each of the characters constitute
the $c=1/2$ Rogers--Ramanujan identities.

For calculations, it is convenient to work with a {\it finitized} partition
function normalized by the largest eigenvalue $T_0(u)$ for which
$\Delta=\overline{\Delta}=k=\overline{k}=0$
\begin{equation}
Z^\sTorus(L;q)=\sum_n
\left(\frac{T_n}{T_{0}}\right)^M \to |q|^{c/12}Z^{\sTorus}(q),\quad 
N,M\to\infty,
\ \mbox{$M/N$ fixed}
\end{equation}
where $n=0,1,\ldots$ labels the eigenvalues $T_n(u)$ of $\T(u)$ for
$N=2L$ columns. The contribution $|q|^{-c/12}$ from the largest
eigenvalue $T_0(u)$ in the limit $M$ and $N$  large with $M/N$ fixed is
obtained in Appendix~B by using the Euler-Maclaurin formula.

Following \cite{Baxter,OPW}, the eigenvalues $T(u)$ of the Ising model
on a torus can be obtained by solving the functional equation
\begin{equation}
T(u) T(u+\case{\pi}{4}) = \cos^{2L}2u-R\,\sin^{2L}2u
\end{equation}
where $L=N/2$,
$T(u)= \overline{T(\case{\pi}{4}-u)}$ and $\T(0)$ is the shift
operator. Here $R=\pm 1$ is the eigenvalue of the spin or height reversal
operator
\begin{equation}
\R_{\smb a,\smb b}=\prod_{j=1}^{N} \delta_{a_j,4-b_j}
\end{equation}
which satisfies $\R^2=\I$, $\R\T(u)=\T(u)\R$.
For $R=1$ the solution of the functional equations is~\cite{OPW}
\begin{equation}
T(u) = \epsilon \sqrt{2} \left(2 e^{2i u+\pi i/4}\right)^{-L}
\prod_{k=1}^L \left( e^{4iu} + i\mu_k
\tan \bigl(\case{(2k-1)\pi}{4L}\bigr)\right)\label{ris1}
\end{equation}
with $\epsilon^2=\mu_k^2=1$ for all $k$ and $\prod_{k=1}^L \mu_k
= 1$. Similarly for $R=-1$, the solution is
\begin{equation}
T(u) = \epsilon \sqrt{L} \left(2 e^{2iu+\pi i/4}\right)^{1-L} \,
\prod_{k=1}^{L-1} \left(e^{4iu} +
i\mu_k \tan \bigl(\case{k\pi}{2L}\bigr)\right)\label{rism1}
\end{equation}
with $\epsilon^2=\mu_k^2=1$ for all $k$.
Up to an overall sign, the eigenvalues are thus determined by the set 
$\{\mu_k\}$.
We note that the total number of eigenvalues is $2^L+2^L=2^{L+1}$ and that
each eigenvalue appears with its negative.

It is often desirable to remove this trivial degeneracy of the eigenvalues $\pm
T(u)$. Following~\cite{OPW} it is convenient to choose
\begin{equation}
\epsilon=\prod_{k=\lfloor (L+3)/2 \rfloor}^L \mu_k,\quad
R=+1;  \qquad \epsilon=\prod_{k=\lfloor (L+2)/2 \rfloor}^{L-1}
\mu_k, \quad R=-1 \label{deg}
\end{equation}
This leads to the {\it finitized} Ising torus
partition function
\cite{OPW}:
\begin{eqnarray}
Z^\sTorus(L;q)&\!\!\!\!=\!\!\!\!&X_0(\lfloor\case{L}{2}\rfloor;\bar{q} 
)\,X_0(\lfloor
\case{L+1}{2}\rfloor;q)+ |q| \, X_{1/2}(\lfloor
\case{L}{2}\rfloor;\bar{q})\, X_{1/2}(\lfloor\case{L+1}{2}\rfloor;q) 
\nonumber\\
& &\qquad\quad +\, |q|^{1/8} \, X_{1/16}(\lfloor\case{L-1}{2}\rfloor;\bar{q})\,
X_{1/{16}}(\lfloor \case{L}{2}\rfloor;q)
\end{eqnarray}
with finitized Virasoro characters~\cite{Melzer}
\begin{eqnarray}
X_0(L;q)&=&\sum_{\{\mu\}_L^+}\prod_{k=1}^L q^{(k-1/2)\delta_{\mu_{k,-1}}}\\
&=&\frac{1}{2}\left\{\prod_{k=1}^L\left(1+q^{k-1/2}\right)+
\prod_{k=1}^L\left(1-q^{k-1/2}\right)\right\}
=\sum_{m=0\atop{m\rm{\ even}}}^L q^{m^2/2} {\Mult{L}{m}}
\label{fchi0}\nonumber\\
X_{1/2}(L;q)&=&q^{-1/2}\sum_{\{\mu\}_L}\prod_{k=1}^L q^{(k-1/2)
\delta_{\mu_{k,-1}}}\\
&=&\frac{1}{2}q^{-1/2}\left\{\prod_{k=1}^L\left(1+q^{k-1/2}\right)-
\prod_{k=1}^L\left(1-q^{k-1/2}\right)\right\}
=\sum_{m=1\atop{m\rm{\ odd}}}^L q^{(m^2-1)/2} {\Mult{L}{m}}
\label{fchihalf}\nonumber\\
X_{1/16}(L;q)&=&\sum_{\{\mu\}_L}\prod_{k=1}^L\left(1-\frac{1}{2}(1-\mu_k)
(1-q^k)\right)\\
&=&\prod_{k=1}^L(1+q^k)
=\sum_{m\geq0}^L q^{m(m+1)/2} {\Mult{L}{m}}.
\label{fchi16}\nonumber
\end{eqnarray}
Here $\{\mu\}_L$ denotes the sequences
$\mu_1,\ldots,\mu_L$ with $\mu_k=\pm1$ and $\{\mu\}_L^{\pm}$
denotes the subset of sequences with $\prod_{k=1}^L\mu_k=\pm1$.
The Gaussian polynomials or $q$-binomials are
\begin{equation}
\Mult{L}{m}=
\sum_{I_1=0}^{L-m} \sum_{I_2=0}^{I_1}\cdots \sum_{I_m=0}^{I_{m-1}}
q^{I_1+\ldots+I_m} =\cases{
\frac{(q)_L}{(q)_{m}(q)_{L-m}}, &\quad $ 0\leq m\leq L $ \cr
0\rule{0in}{.2in}, &\quad otherwise
} \label{Gaussianpoly}
\end{equation}
with $(q)_m=\prod_{n=1}^m (1-q^n)$. The $R=1$ sector of the spectra is given by
$X_0$ and $X_{1/2}$ and the $R=-1$ sector is given by $X_{1/16}$.

\nsection{Ising Model on the Klein Bottle}

\subsection{Flip operator and partition function}\label{fpf}

To form a Klein bottle we consider an $N\times M$ lattice of spins 
$a_{i,j}$ and
impose periodic boundary conditions along the horizontal direction
of the lattice by
identifying column~$N+1$ with column~1
\begin{equation}
a_{i,N+1}=a_{i,1},\qquad i=1,2,\ldots,M.
\end{equation}
Along the vertical direction, we
identify row~$M+1$ with the left-right flip of row~1
\begin{equation}
a_{M+1,j}=a_{1,N+2-j},\qquad j=1,2,\ldots,N.
\label{flipspins}
\end{equation}
This flip reflects about the column $j=1$, or equivalently $j={N\over 2}+1$,
so that the spins at
$j=1$ and $j={N\over 2}+1=L+1$ are left invariant and the even and odd sublattices do not mix under the flip.

The flip operator $\F$ that implements the flip (\ref{flipspins}) has entries
\begin{equation}
\F_{\smb a,\smb b}=\prod_{j=1}^{N} \delta_{a_j,b_{N+2-j}}
\label{flipeqn}
\end{equation}
This matrix has the same dimension as the transfer
matrix $\T(u)$ and satisfies
$\F^2=\I$,
$\F^T=\F$ so that $\F$ has eigenvalues $F=\pm 1$.
Clearly the flip operator $\F$ can be expressed, in a suitable basis, as a
simple block matrix
\begin{equation}
\F=\pmatrix{I_r&0&0\cr 0&0&I_s\cr 0&I_s&0}
\label{directsum}
\end{equation}
where $I_n$ is the $n\times n$ identity matrix, $r$ is the number of row
configurations which are invariant under the flip $\F$, $2s$ is the number of
non-invariant row configurations which are related in pairs by the 
flip $\F$ and
$r+2s=\mbox{dim}\,\F=2(2^{N/2})=2^{L+1}$.

The number of left-right symmetric or invariant row configurations is
simply the number of configurations of length $N/2$ and so
\begin{equation}
r=\sum_{a,b=1}^3 \left[A^{N/2}\right]_{a,b},\qquad\qquad
2s=\mbox{Tr}\,A^{N}-\sum_{a,b=1}^3 \left[A^{N/2}\right]_{a,b}
\end{equation}
Diagonalizing the matrix $A$ and using
spectral decomposition we find that
\begin{equation}
r=(\sqrt{2})^{N\over 2}\left(1+{1\over\sqrt{2}}\right)^2
+(-\sqrt{2})^{N\over 2}\left(1-{1\over\sqrt{2}}\right)^2
=\left\{\begin{array}{ll}
3(2^{L/2}),&\quad \mbox{$L={N\over 2}$ even}\\[6pt]
2^{(L+3)/2},&\quad \mbox{$L={N\over 2}$ odd}
\end{array}
\right. \label{xi}
\end{equation}
We see immediately that $\F$ has $r+s$ eigenvalues $+1$ and $s$ eigenvalues
$-1$ and so $r$ is the net number of positive eigenvalues
\begin{equation}
r=\sum_n F_n.
\end{equation}

The flip operator $\F$ does not commute with the single row transfer matrix
$\T(u)$ so it is not possible to simultaneously diagonalize
$\T(u)$ and
$\F$. Instead, we find
\begin{eqnarray}
\left[\F\T(u)\right]_{\smb a,\smb b}
&\!\!=\!\!&\prod_{j=1}^{N}\W
{b_j}{b_{j+1}}u{a_{N+2-j}}{a_{N+1-j}} \nonumber \\
&\!\!=\!\!&\prod_{j=1}^{N} \left({ S_{a_{N+2-j}}S_{b_{j+1}} \over
S_{a_{N+1-j}}S_{b_j} }\right)^{1/2} \W
{a_{N+2-j}}{a_{N+1-j}}{\lambda-u}{b_j}{b_{j+1}}\\
&\!\!=\!\!&\prod_{j=1}^{N} \W
{a_{N+2-j}}{a_{N+1-j}}{\!\lambda-u}{b_j}{b_{j+1}}\!
= \left[\T(\lambda-u)\F\right]_{\smb a,\smb b}
=\!\Big[\T^T(u)\F\Big]_{\smb a,\smb b}\nonumber
\end{eqnarray}
For this reason we introduce the double row transfer matrix $\D(u)$
defined by
\begin{equation}
\D(u)=\T(\lambda-u)\T(u)=\T^T(u)\T(u)
\end{equation}
Clearly $\D(u)$ is real symmetric and positive definite for real $u$. Moreover,
$\F$ and $\D(u)$ do commute so they can be simultaneously diagonalized
\begin{eqnarray}
\F\D(u)&=&\F\T(\lambda-u)\T(u)=\T(u)\F\T(u)\nonumber\\
&=&\T(u)\T(\lambda-u)\F=\T(\lambda-u)\T(u)\F=\D(u)\F
\end{eqnarray}

The partition function of the Ising model on the Klein bottle is
\begin{equation}
Z^\sKlein_{MN}(u)=\mbox{Tr}\left[\F\D(u)^{M/2}\right]
=\sum_n F_n D_n(u)^{M/2}\label{ptfn}
\end{equation}
where $n$ labels the eigenvalues with respect to the common set
of eigenvectors $\x_n$ of $\F$ and $\D(u)$. Although the flip operator $\F$
reflects about the column $j=1$ we could have chosen a flip
operator
$\F_k$ that reflects about the column $j=k$. In this case
\begin{equation}
\F_k=\C^k\F\C^{-k} \label{flipeqn2}
\end{equation}
It is then easy to check that the partition function \eqref{ptfn} is
translation invariant by using the cyclicity of the trace and the fact that
$\C$ and $\D(u)$ commute
\begin{eqnarray}
\mbox{Tr}\Big[\F_k\D(u)^{M/2}\Big]
&\!\!=\!\!&\mbox{Tr}\Big[\C^k\F\C^{-k}\D(u)^{M/2}\Big]=
\mbox{Tr}\Big[\F\C^{-k}\D(u)^{M/2}\C^k\Big]\nonumber\\
&\!\!=\!\!&\mbox{Tr}\Big[\F\D(u)^{M/2}\Big]
\end{eqnarray}
Hence we only need to work with the flip
operator $\F=\F_1$ defined in \eqref{flipeqn}.

We will show that the simultaneous eigenvectors of
$\F$ and $\D(u)$ are
\begin{equation}
\{\x_n\}=\{\x_1,\ldots,\x_r,\z_1^+,\ldots,\z_s^+,\z_1^-,\ldots,\z_s^-\}
\end{equation}
with
\begin{equation}
\F\x_j=+\x_j,\quad j=1,2,\ldots,r;\qquad
\F\z_k^{\pm}=\pm\z_k^{\pm},\quad k=1,2,\ldots,s
\end{equation}
so that $\F\D(u)^{M/2}\z_k^{\pm}=\pm D(u)^{M/2}\z_k^{\pm}$.
It follows that the eigenvalues corresponding to $\z_k^+$ and 
$\z_k^-$ cancel in
\eqref{ptfn}. So only the eigenvalues of $\D(u)$ corresponding to the even
eigenvectors
\mbox{$\{\x_1,\cdots,\x_r\}$} contribute to the partition function.
We will show that the eigenvectors $\{\x_j\}$ are precisely the real
eigenvectors of $\T(u)$ whereas the eigenvectors $\{\z_k^{\pm}\}$ are
simply related to the complex eigenvectors of $\T(u)$.

\subsection{Simultaneous eigenvectors of $\D(u)$ and $\F$}

In this section we study how the simultaneous eigenvectors of
$\D(u)$ and $\F$ are related to the eigenvectors of $\T(u)$.
Consider the transfer matrix $\T(u)$, with $u$ in the closed interval
$[0,\lambda]\subset{\Bbb R}$. This is a commuting family of
real normal matrices that along with the flip operator $\F$ satisfies
\begin{eqnarray}
&\T(u)\T(v)=\T(v)\T(u),\qquad\T^T(u)=\T(\lambda-u),
\qquad \F\T(\lambda-u)=\T(u)\F&\nonumber\\
&\D(u)=\T^T(u)\T(u),\qquad \F\D(u)=\D(u)\F\qquad \F^2=\I& \label{Fx}
\label{transpose}
\end{eqnarray}

The entries of
$\T(u)$ are real analytic functions of $u$ that can be
analytically continued into the complex plane $u\in{\Bbb C}$.
In general, the eigenvalues $T(u)$ and eigenvectors $\{\z_n\}$ of
$\T(u)$ are complex with the eigenvectors $\z\in\{\z_n\}$
being independent of $u$. Specifically, from (\ref{complexvector}) we have
\begin{equation}
\T(u)\z=T(u)\z,\qquad
\T^T(u)\z=\T(\lambda-u)\z=T(\lambda-u)\z=\overline{T(u)}\z
\end{equation}

The expressions for $T(u)$ can be obtained in the following way.
Given an eigenvector $\z=(z^1,z^2,\ldots)$ with constant complex entries (since
$\z$ is independent of $u$), 
we expand the product $\T(u)\z$ along the row with the largest 
absolute value entry $z^{K}$ in $\z$. 
Since $\z$ is an eigenvector of $\T(u)$, 
the corresponding eigenvalue $T(u)$ is
\begin{equation}
T(u)=\frac{1}{z^K}\sum_j\,[\mathbf{T}(u)]_{K,j}\ z^j
\end{equation}
 Since $\T(u)$ is a finite matrix with entries being analytic functions,
it follows that all the eigenvalues
$T(u)$ which, are the finite sums of analytic functions, are analytic functions of $u$.

We say that two eigenvalues
are degenerate if they agree on the whole interval $[0,\lambda]$.
Clearly, by analytic continuation, two eigenvalues are degenerate
if they agree on any open interval in $[0,\lambda]$. Consequently,
non-degenerate eigenvalues can only cross at isolated points. The
difference between these non-degenerate eigenvalues is an
analytic function with isolated zeros which cannot accumulate in
the closed interval $[0,\lambda]$. It follows that there is only a
finite number of points where the crossings of the various
eigenvalues can occur. Therefore at a ``generic" point $u=u_0$
there are no accidental crossings, that is, any two eigenvalues
are either degenerate on $[0,\lambda]$ or distinct at $u=u_0$. We
conclude that the simultaneous eigenvectors of $\T(u)$ are
uniquely determined (up to linear combinations for the degenerate
eigenvalues) by the distinct eigenvectors at $u=u_0$. So, in
practice numerically, it is sufficient to fix $u$ to a ``generic"
value $u_0$ and diagonalize $T(u_0)$ to find the simultaneous
eigenvectors.

Let $\{\z_n\}$ be a set of orthonomal complex eignvectors of $\T(u)$.
We want to construct a set of simultaneous eigenvectors $\{\x_n\}$ of
$\D(u)=\T^T(u)\T(u)$ and $\F$. Suppose, after suitable
normalization, that the eigenvector $\z=\x$ of $\T(u)$ is real. Then
it follows that the corresponding eigenvalue $T(u)$ is real for
all real $u$. We refer to these as the real eigenvalues of
$\T(u)$.
\begin{equation}
\T(u)\x_j=\T^T(u)\x_j=T(u)\x_j,\qquad \D(u)\x_j=T(u)^2\x_j,
\qquad j=1,2,\ldots,r
\end{equation}
In the remaining case we have to work with complex
eigenvectors with both real and imaginary parts non-zero. In this
case, the eigenvalues appear in complex conjugate pairs since
$\z$ and $\overline{\z}$ are linearly independent
and $\T(u)\z=T(u)\z$ implies
\begin{equation}
\T(u)\overline{\z}=\overline{\T(u)\z}=\overline{T(u)\z}=\overline{T(u)
}\overline{\z}.
\end{equation}
Hence
\begin{equation}
\T(u)\z_k=T(u)\z_k,\quad \T^T(u)\z_k=\overline{T(u)}\z_k,\quad
\D(u)\z_k=|T(u)|^2\z_k,\quad k=1,2,\ldots,s
\end{equation}
with a similar statement for $\overline{\z}_k$.

Let $\{\x_1,\ldots,\x_r,\z_1,\ldots,\z_s\}$ be the set of mutually
orthonormal real and complex eigenvectors corresponding to the real
and complex eigenvalues
$T(u_0)\in{\Bbb R}$, $T(u_0)\in{\Bbb C}$
with $\Im T(u_0)>0$. Then the $r+2s$ mutually orthonormal
simultaneous eigenvectors of $\D(u)=\T^T(u)\T(u)$ and $\F$ are
\begin{equation}
\{\x_1,\ldots,\x_r,\x_1^+,\ldots,\x_s^+,\x_1^-,\ldots,\x_s^-\}\label{orthoset}
\end{equation}
where
\begin{equation}
\F\x_j=F\x_j,\qquad \x_k^+=\z_k^+,\qquad \x_k^-=-i\z_k^-
\qquad\z_k^\pm={1\over\sqrt{2}}(\I\pm \F)\z_k
\end{equation}
and the corresponding eigenvalues of $\{\T(\lambda-u)\T(u),\F\}$
are $\{T(u)^2,F\}$, $\{|T(u)|^2,+1\}$ and $\{|T(u)|^2,-1\}$
respectively for the three groups. Although it is not apparent here we will see
later that, since
$\F\z_k=\overline{\z}_k$, the vectors $\x_j,\x_k^+,\x_k^-$ are all real with
$\x_k^+=\sqrt{2}\Re(\z_k)$, $\x_k^-=\sqrt{2}\Im(\z_k)$.

We show that (\ref{orthoset}) is an
orthonormal set of eigenvectors and hence a complete set of linearly
independent eigenvectors of $\T(\lambda-u)\T(u)$ and $\F$.
To establish this we note some properties of $\T(u)$
and $\F$ (see Appendix~A):
\begin{itemize}
\item Two eigenvectors $\z_i$ and $\z_j$ of $\T(u)$ are orthogonal if
their eigenvalues are distinct at $u=u_0$.
\item If $\T(u)\z=T(u)\z$ then $\F\z$ is also an
eigenvector of $\T(u)$ with the corresponding eigenvalue
$\overline{T(u)}$ since
\begin{equation}
\T(u)\left(\F\x\right)=(\T(u)\F)\x=\left(\F\T(\lambda-u)\right)\x
      =\overline{T(u)}\F\x\label{cconj}
\end{equation}
\end{itemize}
Now we observe that the eigenvalues corresponding to the eigenvectors
$\z_k$ and $\F\z_\ell$ are distinct since they have imaginary
parts $\Im T(u)>0$ and $\Im \overline{T(u)}<0$ respectively. So
these two eigenvectors are orthogonal
\begin{equation}
\z_k^\dagger\F\z_\ell=0,\quad k,\ell=1,2,\ldots,s
\end{equation}
The required orthonormality now follows by straigthforward
calculation of inner products
\begin{eqnarray}
(\z_k^\pm)^\dagger\x_j
&=&{1\over\sqrt{2}}\,\z_k^\dagger(\I\pm\F)\x_j
={1\over\sqrt{2}}(1\pm F)\;\z_k^\dagger\x_j=0\\
(\z_k^\pm)^\dagger\z_\ell^\pm
&=&{1\over 2}\,\z_k^\dagger(\I\pm\F)^2\z_\ell
=\z_k^\dagger(\I\pm\F)\z_\ell=\z_k^\dagger\z_\ell=\delta_{k,\ell}\\
(\z_k^\pm)^\dagger\z_\ell^\mp
&=&{1\over 2}\,\z_k^\dagger(\I\pm\F)(\I\mp\F)\z_\ell=0
\end{eqnarray}

Similarly, the required eigenvalues follow by straightforward
calculation
\begin{eqnarray}
\T(\lambda-u)\T(u)\x_j&=&\overline{T(u)}T(u)\x_j=T(u)^2\x_j,\qquad
\F\x_j=F\x_j\\
\T(\lambda-u)\T(u)\z_k^\pm
&=&{1\over\sqrt{2}}\,\T(\lambda-u)\T(u)(\I\pm \F)\z_k\nonumber\\
&=&{1\over\sqrt{2}}\,\T(\lambda-u)\left(T(u)\z_k\pm\overline{T(u)}\F\z
_k\right)\\
&=&{1\over\sqrt{2}}\,\left(|T(u)|^2\z_k\pm |T(u)|^2\F\z_k\right)
=|T(u)|^2\z_k^\pm\nonumber\\
\F\z_k^\pm&=&{1\over\sqrt{2}}\,\F(\I\pm\F)\z_k=\pm
{1\over\sqrt{2}}\,(\I\pm\F)\z_k=\pm\z_k^\pm
\end{eqnarray}

\subsection{Eigenvectors of $\T(u)$}

To study the simultaneous eigenvectors of $\T(u)$ and $\T^T(u)$, we fix
$u=u_0$ to a generic value to avoid accidental crossings of eigenvalues.
We also set $\T=\T(u_0)$ and $T=T(u_0)$ so that the eigenvector
equations for the common eigenvectors are
\begin{equation}
\T\z=T\z,\qquad \T^T\z=\overline{T}\z,\qquad \D\z=\T^T\T\z=|T|^2\z
\label{eigvec}
\end{equation}
The transfer matrix $\T$ does not commute with the flip operator $\F$ but
the double row transfer matrix $\D=\T^T\T$ does.
In this section we obtain the structure of the complex eigenvectors $\z$.

Let us order the $r+2s$ paths $\a$
into three blocks of size $r$, $s$ and $s$ respectively. In the first block
we put the $r$ paths that are invariant under the flip $\F$. The remaining
$2s$ paths occur in pairs related by the flip and we place these in order
in the second and third blocks. In this basis, the block structure of
$\F$, $\T$ and $\T^T$ is
\begin{equation}
\F=\pmatrix{I&0&0\cr 0&0&I\cr 0&I&0},\qquad
\T=\pmatrix{A&B&C\cr C^T&D&H\cr B^T&G&D^T},\qquad
\T^T=\pmatrix{A&C&B\cr B^T&D^T&G\cr C^T&H&D}
\end{equation}
where $A^T=A$, $H^T=H$ and $G^T=G$ and we have used the fact that
$\F\T=\T^T\F$.

The $r+s$ even and $s$ odd real eigenvectors of $\F$ are
of the form
\begin{equation}
\pmatrix{\a\cr \b\cr \b},\qquad \pmatrix{0\cr \c\cr -\c}
\label{evodd}
\end{equation}
It follows that the $r+s$ even and $s$ odd common eigenvectors
of
$\D=\T^T\T$ and $\F$ are real linear combinations of these $r+s$
even and $s$ odd eigenvectors respectively.
Since the real symmetric matrices $\D=\T^T\T$ and $\F$ commute it
follows that the real and imaginary parts of $\z$ in (\ref{eigvec}) are
common real eigenvectors of $\D$ and $\F$ of the form(\ref{evodd}).
Consequently, there are three possibilities (i)
$\z$ is even and $\F\z=\z$, (ii) $\z$ is odd and $\F\z=-\z$ or (iii)
$\z$ is mixed so that after normalization $\F\z=\overline{\z}$ 
(In fact, it is not hard to show that $\F\z=e^{i\theta}\overline{\z}$
 for any arbitrary real $\theta$ if $\z$ is mixed). But
\begin{eqnarray}
\T\z=T\z&\Rightarrow&
\T^T\z=\overline{T}\z\quad\mbox{and}\quad \T^T(\F\z)=T(\F\z)\nonumber\\
&\Rightarrow& (T-\overline{T})\z^\dagger(\F\z)=0
\;\Rightarrow\; \Im(T)\,\z^\dagger(\F\z)=0
\end{eqnarray}
So if $T\notin {\Bbb R}$, we conclude that $\z$ and $\F\z$ are orthogonal.
Hence $\F\z\ne\pm\z$ and consequently $\F\z=\overline{\z}$ and
the complex eigenvectors of $\T$ occur in complex conjugate pairs
\begin{equation}
\z=\pmatrix{\a\cr \b+i\c\cr \b-i\c},\qquad
\F\z=\overline{\z}=\pmatrix{\a\cr \b-i\c\cr \b+i\c}
\label{complexvec}
\end{equation}

We can write the eigenvector equations for $\T$ and $\T^T$ in terms of the
symmetric and skew-symmetric combinations
\begin{eqnarray}
(\T+\T^T)\z&=&\pmatrix{
2A&B+C&B+C\cr B^T+C^T&D+D^T&H+G\cr B^T+C^T&H+G&D+D^T}\z
=2\Re(T)\z \label{symmskew1} \\
(\T-\T^T)\z&=&\pmatrix{
0&B-C&C-B\cr C^T-B^T&D-D^T&H-G\cr B^T-C^T&G-H&D^T-D}\z
=2i\Im (T)\z
\label{symmskew}
\end{eqnarray}
We now look for a complex eigenvector of the form (\ref{complexvec}) and
take the real and imaginary parts in (\ref{symmskew1},\ref{symmskew})
\begin{eqnarray}
\pmatrix{
2A&B+C&B+C\cr B^T+C^T&D+D^T&H+G\cr B^T+C^T&H+G&D+D^T}
\pmatrix{\a\cr \b\cr \b}
&=&2\Re(T)\pmatrix{\a\cr \b\cr \b}\\[6pt]
\pmatrix{
2A&B+C&B+C\cr B^T+C^T&D+D^T&H+G\cr B^T+C^T&H+G&D+D^T}
\pmatrix{0\cr \c\cr -\c}
&=&2\Re(T)\pmatrix{0\cr \c\cr -\c}\\[6pt]
\pmatrix{
0&B-C&C-B\cr C^T-B^T&D-D^T&H-G\cr B^T-C^T&G-H&D^T-D}
\pmatrix{\a\cr \b\cr \b}
&=&-2\Im (T)\pmatrix{0\cr \c\cr -\c}\\[6pt]
\pmatrix{
0&B-C&C-B\cr C^T-B^T&D-D^T&H-G\cr B^T-C^T&G-H&D^T-D}
\pmatrix{0\cr \c\cr -\c}
&=&2\Im (T)\pmatrix{\a\cr \b\cr \b}\label{skewab}
\end{eqnarray}
The symmetric equations can be simplified and from the two skew-symmetric
equations we obtain two eigenvector equations for real symmetric matrices
\begin{eqnarray}
\pmatrix{
2A&\sqrt{2}(B+C)\cr \sqrt{2}(B+C)^T&D+D^T+H+G}
\pmatrix{\a\cr \sqrt{2}\b}
&=&2\Re(T)\pmatrix{\a\cr \sqrt{2}\b}\label{abeqn1}\\[6pt]
\pmatrix{
D+D^T&H+G\cr H+G&D+D^T}
\pmatrix{\c\cr -\c}
&=&2\Re(T)\pmatrix{\c\cr -\c}\label{ceqn1}\\[6pt]
\pmatrix{
0&B-C&C-B\cr C^T-B^T&D-D^T&H-G\cr B^T-C^T&G-H&D^T-D}^2
\pmatrix{\a\cr \b\cr \b}
&=&-4(\Im\, T)^2\pmatrix{\a\cr \b\cr \b}\label{abeqn2}\\[4pt]
\pmatrix{
0&B-C&C-B\cr C^T-B^T&D-D^T&H-G\cr B^T-C^T&G-H&D^T-D}^2
\pmatrix{0\cr \c\cr -\c}
&=&-4(\Im\, T)^2\pmatrix{0\cr \c\cr -\c}
\label{ceqn2}
\end{eqnarray}

Equation (\ref{ceqn2}) now reduces to a simple eigenvector
equation for a real symmetric matrix
\begin{equation}
[2(B-C)^T(B-C)-(D-D^T)^2+(H-G)^2]\c=4(\Im\,T)^2\c
\end{equation}
From numerical observation, this equations yields $s$ real eigenvectors with
$\Im\,T\ne 0$ at any generic point $u$. This is verified in section 
\ref{finiteklein} 
by counting the number of real eigenvalues of $\T$ which matches the net number of positive eigenvalues of $\F$ in \eqref{xi}.
Indeed, we will see that this generic situation leads to $s$
complex conjugate pairs of eigenvectors of the form (\ref{complexvec}).
A non-generic choice of $u$
would lead to fewer than $s$ complex conjugate eigenvalues and
eigenvectors. 
From (\ref{ceqn1}) we see that, in the generic situation,
the $s$ vectors $\c$ must also satisfy the simple real symmetric
eigenvector equation
\begin{equation}
[D+D^T-H-G]\c=2\Re(T)\c
\end{equation}
For a given solution vector $\c$ we have thus determined the real and
imaginary parts of the eigenvalues $T=\Re(T)\pm i\Im(T)$. The
vectors $\a$ and $\b$ are determined by (\ref{skewab})
\begin{equation}
\a=(\Im\,T)^{-1}(B-C)\c,\qquad
\b=(2\Im\,T)^{-1}(D-D^T-H+G)\c,\qquad \Im(T)\ne 0
\end{equation}
We have thus completely determined the $s$ pairs of complex eigenvectors
$\z,\overline{\z}$ of (\ref{complexvec}) corresponding to the
complex eigenvalues $T=\Re(T)\pm i\Im(T)$ with $\Im(T)\ne 0$. The remaining
equations such as (\ref{abeqn1}) that we did not use in deriving the these
eigenvectors must be automatically satisfied.

Each of the $s$ complex conjugate pairs of eigenvectors gives rise to one
even eigenvector $\x^+=\sqrt{2}\Re(\z)$ and one odd eigenvector
$\x^-=\sqrt{2}\Im(\z)$ of
$\F$ and $\D=\T^T\T$. These completely exhaust the odd common eigenvectors
of $\F$ and $\D=\T^T\T$. In the generic situation, it therefore remains to
obtain the remaining $r$ real even eigenvectors of $\T$
\begin{equation}
\x=\pmatrix{\a\cr \b\cr \b}
\end{equation}
which necessarily satisfy
\begin{equation}
\F\x=\x,\qquad \T\x=\T^T\x=T\x,\qquad \D\x=\T^T\T\x=|T|^2\x,\qquad
T\in{\Bbb R}
\end{equation}
These eigenvectors are in fact given by the symmetric eigenvector equation
(\ref{abeqn1}) which yields $r+s$ even eigenvectors. The $r$ solutions that
we need are the solutions that remain after the $s$ solutions for $(\a,\b)$
obtained in the complex case are removed.

In a non-generic situation, such as $u=0$ when $\T$ reduces to the shift
operator, there are fewer than $s$ complex conjugate pairs of eigenvectors
of $\T$. In this case, each complex conjugate pair that is lost is replaced
by one even and one odd real eigenvector. This is the only time that odd
real eigenvectors of $\T$ occur.
We therefore conclude that, in the generic situation, all eigenvectors $\z$
of $\T$ satisfy
\begin{equation}
\F\z=\overline{\z}
\end{equation}
In this case the common real eigenvectors of
$\D=\T^T\T$ and $\F$ are
\begin{eqnarray}
\x^+&=&{1\over\sqrt{2}}(\I+\F)\z=\sqrt{2}\pmatrix{\a\cr \b\cr\b}
=\sqrt{2}\Re(\z)\\
\x^-&=&-{i\over\sqrt{2}}(\I-\F)\z=\sqrt{2}\pmatrix{0\cr \c\cr -\c}
=\sqrt{2}\Im(\z)
\end{eqnarray}
Clearly, these eigenvectors are of the form
(\ref{evodd}).

\subsection{Conformal finite-size corrections}

Consider the finite-size partition function $Z_{NM}$ of the
critical Ising model on a Klein bottle formed from a lattice with $N$ 
columns and
$M$ rows. The asymptotic behaviour of  $Z_{NM}$  in the
limit of large $N$ and $M$ with the aspect ratio $M/N$ fixed is given by
\begin{equation}
Z^\sKlein_{NM}(u)=\mbox{Tr} [\F\D(u)^{M/2}]\sim\exp\left[-NM
\fb(u)-M\fs(u)\right]\, Z^\sKlein(q) \label{ZNM2}
\end{equation}
where $\D(u)$ is the double row transfer matrix, $\fb(u)$ is the bulk 
free energy,
$\fs(u)$ is the boundary free energy and $Z^\sKlein(q)$ is the 
universal conformal
partition function. Since we are working with a double row such that 
$D(u)=T(u)\,T(\lambda-u)$ and that $q_\sTorus(\lambda-u)=\overline{q_\sTorus(u)}$, 
the product $q_\sTorus(u)\,\overline{q_\sTorus(u)}$
is real positive and thus we define 
the modular
parameter to be real and given by
\begin{equation}
q = \exp\!\Big(\!-2\pi\,{M\over N} \sin 4u\Big)
=|q_\sTorus|\label{modklein}
\end{equation}
where $q_\sTorus$ is given by (\ref{modtor}).
Because of the presence of the flip, the left and right
movers mix. Consequently, as is the case on the cylinder, there is 
only one copy
of the Virasoro algebra. 
The leading corrections to each transfer 
matrix eigenvalue
$D(u)$ are 
obtained from the torus case 
by identifying 
$\Delta$ with $\overline{\Delta}$ 
and $k$ with $\overline{k}$
in \eqref{pcorr}
\begin{equation}
\frac{1}{2} \log D(u) = -N\fb(u)-\fs(u)+\frac{4\pi}{N}
\left(\frac{c}{24}-\Delta-k\right)\sin(4u)+
\mbox{o}\left(\frac{1}{N}\right) \label{scal}
\end{equation}
with $\Delta=0,1/2,1/16$ and $k$ a non-negative
integer. 
Cardy \cite{Cardy} argued that as the model is conformal invariant, all spectra are organized into Virasoro characters, and thus
the eigenvalues naturally divide into three towers with
\begin{equation}
Z^\sKlein(q) = \sum_{\Delta} {\cal N}(\Delta)\; (q^2)^{-c/24+\Delta} \sum_k
q^{2k}= \sum_{\Delta} {\cal N}(\Delta) \chi_{\Delta}(q^2)
\label{Zq}
\end{equation}
where ${\cal N}(\Delta)\in \{0,1,2,\ldots\}$  is the operator content on the Klein bottle. Equation \eqref{Zq} with ${\cal N}(\Delta)=1$ is confirmed numerically in the next section.

\subsection{Partition function of the Ising model on the Klein
Bottle}\label{finiteklein}

The {\it finitized} conformal partition function on the Klein bottle is
\begin{equation}
Z^\sKlein(L;q)=\sum_n F_n
\left(\frac{D_n}{D_{0}}\right)^{M/2}
\end{equation}
where $n=0,1,\ldots$ labels the eigenvalues $D_n$ of $\D(u)$ and 
$F_n$ of $\F$ for
$N=2L$ columns and $D_{0}$ is the largest eigenvalue.
We will obtain this partition function from the eigenvalues $T(u)$ of the
periodic row transfer matrix as given by \eqref{ris1}, \eqref{rism1}. 
To do this
we need to separate out the $r$ real eigenvalues from the
$2s$ eigenvalues which appear in complex conjugate pairs.

It can be shown~\cite{Baxter} that the eigenvalues $T(u)$ possess the 
periodicity
\begin{equation}
T(u+\frac{\pi}{2})=(-1)^L\,T(u)
\end{equation}
and that within the period strip $-\frac{\pi}{8}\le 
\Re(u)<\frac{7\pi}{8}$ in the
complex $u$-plane the eigenvalues have exactly $2P$ zeros where
\begin{equation}
P=\cases{L,&$R=+1$\cr L-1,&$R=-1$}
\end{equation}
For the ground state, given by $\mu_k=1$ for all
$k$, the zeros occur within the period strip as 2-strings on the lines
$\Re(u)=-\frac{\pi}{8},\frac{3\pi}{8}$. For excitations, zeros also 
occur on the
lines $\Re(u)=\frac{\pi}{8},\frac{5\pi}{8}$.
The eigenvalues are in fact completely determined by the patterns of 
zeros in the
complex $u$-plane. Indeed we will show that $T(u)$ can be uniquely
expressed in the form
\begin{equation}
T(u)=R^\pm \prod_{k=1}^P \sin2(u-u_k),\qquad
R^\pm,u_1,\cdots,u_P\in\Bbb C \label{eigenexpress}
\end{equation}
and obtain explicit expressions for the constant $R^\pm$ and zeros $u_k$.

Let us define the monotonically increasing sequences
\begin{equation}
t_k=\cases{
      \tan\left({(2k-1)\pi\over 4L}\right), &\quad $R=+1,\quad k=1,\ldots,L$\cr
      \tan({k\pi\over 2L}), \rule{0in}{.23in}&\quad $R=-1,\quad k=1,\ldots,L-1$}
\end{equation}
so that
\begin{equation}
t_k=t_{P+1-k}^{-1},\qquad k=1,\ldots,P
\end{equation}
\begin{equation}
t_{\lceil\frac{P}{2}\rceil}=1, \quad \mbox{$P$ odd};\qquad\qquad
\prod_{k=1}^P t_k=1.\label{prodt}
\end{equation}

For $R=+1$, we can now re-express \eqref{ris1} in the form
\begin{eqnarray}
T(u) &=& \epsilon \sqrt{2} \left(2 e^{2i u+\frac{\pi i}{4}}\right)^{-L}
\prod_{k=1}^L \left( e^{4iu} + i\mu_k
\tan \bigl(\case{(2k-1)\pi}{4L}\bigr)\right)\nonumber\\
&=&\epsilon\sqrt{2}
\prod_{k=1}^L\frac{1}{2i}\left[e^{2i u+\frac{\pi i}{4}}+t_k
e^{\frac{\pi i}{2}\mu_k-2i u+\frac{\pi i}{4}}\right]
\label{zeroplus1}\\
&=&\epsilon\sqrt{2}\prod_{k=1}^L
t_k^{1\over 2}e^{\frac{\pi i}{4}(\mu_k-1)}\frac{1}{2i}
\left[e^{2i\left(u+\frac{i}{4}\log
t_k-\frac{\pi}{8}\mu_k+\frac{\pi}{4}\right)}-e^{-2i\left(u+\frac{i}{4}\log
t_k-\frac{\pi}{8}\mu_k+\frac{\pi}{4}\right)}\right]\nonumber \\
&=&\epsilon\sqrt{2}
e^{\frac{\pi i}{4}\left[-L+\sum_{k=1}^L
\mu_k\right]}\prod_{k=1}^L
\sin2\Big({\textstyle u+\frac{i}{4}\log
t_k-\frac{\pi}{8}\mu_k+\frac{\pi}{4}}\Big)
\nonumber\\
&=&R^+ \prod_{k=1}^P \sin2(u-u_k)\nonumber
\end{eqnarray}
where $\epsilon^2=\mu_k^2=\prod_{k=1}^L \mu_k=1$ and we have used
(\ref{prodt}).  Suppose, for a given eigenvalue $T(u)$, that $m$ is the
number of
$\mu_k$ such that
$\mu_k=-1$. Since $\prod_{k=1}^L\mu_k=1$, $m=2\rho$ is even and
\begin{equation}
\sum_{k=1}^L\mu_k=-m +(L-m)=L-2m=L-4\rho
\end{equation}
Hence
\begin{equation}
e^{\frac{\pi i}{4}\left(-L+\sum_{k=1}^L\mu_k\right)}=e^{-\pi i\rho}=
(-1)^{\rho}
\end{equation}
so that \eqref{zeroplus1} becomes
\begin{equation}
T(u)=(-1)^\rho \epsilon\,\sqrt{2}
\left[\sin2(u-u_{\lceil\frac{L}{2}\rceil})\right]^{\delta_{
L\bmod 2,1}}
\prod_{k=1}^{\lfloor L/2\rfloor}\!\!
\sin2(u-u_k^+)\sin2(u-u_k^-)\label{zeroplus2}
\end{equation}
where
\begin{eqnarray}
u_k&=&\frac{\pi}{8}\mu_k-\frac{\pi}{4}-\frac{i}{4}\log t_k,\qquad
k=1,2,\ldots,P\\
u_k^+&=&u_k=\frac{\pi}{8}\mu_k-\frac{\pi}{4}-\frac{i}{4}\log t_k,\qquad
k=1,2,\ldots,\lfloor P/2\rfloor\\
u_k^-&=&u_{L+1-k}=\frac{\pi}{8}\bar{\mu}_k-\frac{\pi}{4}+\frac{i}{4}
\log t_k,\qquad
k=1,2,\ldots,\lfloor P/2\rfloor
\end{eqnarray}
and
\begin{equation}
\bar{\mu}_k=\mu_{P+1-k},\qquad
k=1,\ldots,\lfloor P/2\rfloor.
\end{equation}
Here $u_k^+$, $u_k^-$ are the zeros in the upper and lower half planes
respectively. Note that, if $\mu_k=\bar\mu_k$, then  $u_k^-$ is simply the
complex conjugate of $u_k^+$. Note also that real zeros $u_k\in{\Bbb R}$
occur for
\begin{equation}
k=
\lceil P/2\rceil, \qquad\mbox{$P$ odd}
\end{equation}

Similarly, for $R=-1$, \eqref{rism1} becomes
\begin{eqnarray}
T(u) &=& \epsilon \sqrt{L} \left(2 \e^{2iu+\frac{\pi i}{4}}\right)^{1-L} \,
\prod_{k=1}^{L-1} \left( \e^{4iu} +
i\mu_k \tan \bigl(\case{k\pi}{2L}\bigr)\right)\nonumber\\
&=&\epsilon\,\sqrt{L}\,\e^{\frac{\pi i}{4}(1-L)}
\prod_{k=1}^{L-1}\frac{1}{2}\left[e^{2i u}+t_k e^{-2i u+
\frac{\pi i}{2}\mu_k}\right]
\label{zerominus1}\\
&=&\epsilon\,\sqrt{L}\,e^{\frac{\pi i}{4}(1-L)}e^{\frac{\pi i}{4}
\sum_{k=1}^{L-1}\mu_k}\prod_{k=1}^{L-1}
\sin2\Big({\textstyle u+\frac{i}{4}\log t_k-\frac{\pi}{8}\mu_k+
\frac{\pi}{4}}\Big)\nonumber\\
&=&R^- \prod_{k=1}^P \sin2(u-u_k)\nonumber
\end{eqnarray}
where $\epsilon^2=\mu_k^2=1$ for all $k$ and we have again used (\ref{prodt}).
But now
\begin{equation}
\exp\!\left[\frac{\pi i}{4}\Big(1-L+\sum_{k=1}^{L-1}\mu_k\Big)\right]
=\exp\!\Big(\!-\frac{\pi i}{4}\,2m\Big)=(-i)^m
\end{equation}
where $m$ is the number of $\mu_k$ such that
$\mu_k=-1$. Hence \eqref{zerominus1} reduces to
\begin{equation}
T(u)=\epsilon\,(-i)^m\,\sqrt{L}
\left[\sin2(u-u_{\lceil\frac{L-1}{2}\rceil})\right]^{\delta_{L
\bmod 2,0}}
\prod_{k=1}^{\lfloor\frac{L-1}{2}\rfloor}\!\!
\sin2(u-u_k^+)\sin2(u-u_k^-)
  \label{zerominus2}
\end{equation}

We want to pick out the eigenvalues of $\T(u)$ which are
real for all $u\in\Bbb R$. First, suppose
the zeros occur in complex conjugate pairs
and $R^\pm$ is real. Then it is clear that $T(u)$ is real for all
$u\in\Bbb R$.
Conversely, suppose
$T(u)$ is real
for all $u\in\Bbb R$. Then
\begin{equation}
\overline{T(u)}=\overline{R^\pm}\prod_{k=1}^P\sin 2(u-\overline{u}_k)
=T(u)\, ,\qquad u\in\Bbb R
\end{equation}
which implies that the zeros $u_k$ occur in complex conjugate
pairs and that $R^\pm$ is real.
Thus, $T(u)$ is real for all $u\in\Bbb R$ if and only if $R^\pm$ is 
real and the
zeros $u_k$ occur in complex conjugate pairs, that is, if the 
patterns of zeros in
the upper and lower half planes are related by complex conjugation. Since
$\Delta$ and $k$ are determined by the pattern of zeros in the upper half plane
and $\overline{\Delta}$ and $\overline{k}$ are determined by the 
pattern of zeros
in the lower half plane, this picks out precisely the eigenvalues for which
$\Delta=\overline{\Delta}$ and $k=\overline{k}$ \cite{Klumper}.
By \eqref{zeroplus2} and \eqref{zerominus2}, we see that the eigenvalue
$T(u)$ is real for all $u\in\Bbb R$ if and only if
\begin{equation}
\mu_k=\overline{\mu}_k \mbox{ and } \prod\limits_{k=1}^{P}\mu_k=1.
\label{realcondition}
\end{equation}
For $R=-1$ and $L$ even ($P$ odd), the last condition in
\eqref{realcondition} reduces to
$\mu_{\lceil\frac{L-1}{2}\rceil}=1$.

Using \eqref{realcondition} and allowing for $\epsilon=\pm1$, we can count
the number $r$ of real eigenvalues $T(u)\in\Bbb R$
\begin{equation}
r=2\Big(2^{\lfloor\frac{L}{2}\rfloor}+2^{\lfloor\frac{L-1}{2}\rfloor}\Big)
=\left\{\begin{array}{ll}
3(2^{L/2}), & \qquad\mbox{$L$ even} \\
2^{(L+3)/2}, & \qquad\mbox{$L$ odd}
\end{array}\right.\label{effdim}
\end{equation}
in agreement with \eqref{xi}. Since $r$ is the net number of positive
eigenvalues of the flip operator $\F$, as discussed in Section~3.1, any real
eigenvector of $\T(u)$ is an eigenvector of $\F$ with $F=+1$. The 
remaining $2s$
eigenvectors occur in complex conjugate pairs with $F=\pm 1$.
After cancellation of the contributions from the complex eigenvalues 
and removal
of the trivial $\pm T(u)$ degeneracy, the partition function for
the Ising model on the Klein bottle is
\begin{equation}
Z^\sKlein_{NM}(u)=\sum_n F_n D_n(u)^{M/2}
=\sum_{n=0}^{r/2-1}  D_n(u)^{M/2}=\sum_{n=0}^{r/2-1} |T_n(u)|^M
\end{equation}
where the sum is over the $r/2$ real eigenvalues of $\D(u)$
\begin{equation}
D_n(u)=|T_n(u)|^2,\qquad n=0,1,\ldots,r-1,\quad T_n(u)\in {\Bbb R}
\end{equation}

\subsection{Finitized Ising partition function on Klein
bottle}\label{characters}

In this section, we follow \cite{OPW} to obtain the {\it finitized} conformal
partition function
$Z^{\sKlein}(L;q)$ in terms of finitized Virasoro characters.

We begin by using (\ref{deg}) to remove the degeneracy in the eigenvalue
expressions (\ref{ris1}) and (\ref{rism1}) for the single row transfer matrix
eigenvalues $T(u)$.  This degeneracy is irrelevant for the eigenvalues
$D(u)=\overline{T(u)}T(u)$ of the double row transfer matrix. Using 
(\ref{deg}) and
\eqref{ris1} we find for $R=+1$
\begin{eqnarray}
D(u)&=&\overline{T(u)}T(u) = 2^{-2L+1}
\left|e^{4iu}+i\mu_{\lceil\frac{L}{2}\rceil}
\tan\bigl(\case{\pi(2\lceil L/2\rceil-1)}{4L}\bigr)
\right|^{2\delta_{L\bmod2,1}} \nonumber \\
&&\mbox{}\hspace{-.5in}\times
\prod_{k=1}^{\lfloor L/2\rfloor}
\left|e^{4iu} + i\mu_k \tan
\bigl(\case{\pi(2k-1)}{4L}\bigr)\right|^2
\left|\bar{\mu}_k e^{4iu} +
i \cot \bigl(\case{\pi(2k-1)}{4L}\bigr)\right|^2
\label{fac1}
\end{eqnarray}
with $\bar{\mu}_k = \mu_{L-k+1}$. Similarly, from (\ref{rism1})
we find for $R=-1$
\begin{eqnarray}
D(u)&=&\overline{T(u)}T(u) = 2^{2(1-L)} L
\left|e^{4iu}
+ i\mu_{\lceil\frac{L-1}{2}\rceil} \tan \bigl(\case{\pi
\lceil\frac{L-1}{2}\rceil}{2L}\bigr)
\right|^{2\delta_{L\bmod 2,0}} \nonumber\\
&&\mbox{}\hspace{-.5in}\times
\prod_{k=1}^{\lfloor
\frac{L-1}{2} \rfloor} \left|e^{4iu} + i\mu_k \tan
\bigl(\case{\pi k}{2L}\bigr)\right|^2
\left|\bar{\mu}_k
e^{4iu} + i \cot \bigl(\case{\pi k}{2L}\bigr)\right|^2
\label{fac2}
\end{eqnarray}
with $\bar{\mu}_k = \mu_{L-k}$. We remark that although the forms for 
$T(u)$ are
not unique due to the choice of $\epsilon$, the forms given here for $D(u)$ are
unique since they are independent of $\epsilon$.

Taking the ratio of (\ref{fac1}) and (\ref{fac2}) with the
largest eigenvalue $D_0(u)$ given by (\ref{fac1}) with all
$\mu_k=\bar{\mu}_k=1$ and using the limits
\begin{eqnarray}
\lim_{N\to\infty} N \log\!\left[\frac{e^{4i
u}-i\tan\bigl(\frac{\pi K}{N}\bigr)} {e^{4i
u}+i\tan\bigl(\frac{\pi K}{N}\bigr)}\right]
&=& -2\pi i K \e^{-4iu} \nonumber \\
\lim_{N\to\infty} N \log\!\left[\frac{-e^{4i
u}+i\cot\bigl(\frac{\pi K}{N}\bigr)} {e^{4i
u}+i\cot\bigl(\frac{\pi K}{N}\bigl)}\right] &=& 2\pi i K
\e^{4iu}
\end{eqnarray}
along with the condition $\eqref{realcondition}$ and the definition
of $q$ in $\eqref{modklein}$, we obtain
\begin{eqnarray}
Z^\sKlein(L;q) &=& \sum_{\{\mu\}_{\lfloor L/2\rfloor}^+}
\,\prod_{k=1}^{\lfloor L/2\rfloor}
q^{2(k-1/2)\delta_{\mu_k,-1}}+ \sum_{\{\mu\}_{\lfloor
L/2\rfloor}^-} \,\prod_{k=1}^{\lfloor L/2\rfloor}
q^{2(k-1/2)\delta_{\mu_k,-1}}
\nonumber \\
& & \qquad +|q|^{1/8}\hspace{-3mm} \sum_{\quad \{\mu\}_{\lfloor
(L-1)/2\rfloor}} \hspace{-3mm} \prod_{k=1}^{\lfloor
(L-1)/2\rfloor}q^{2k\delta_{\mu_k,-1}}
\end{eqnarray}
Here the factor $|q|^{1/8}$ arises from the scaling limit of the
ratio of the largest eigenvalue in (\ref{fac2}), which has all
$\mu_k=\bar{\mu}_k=1$, to the overall largest eigenvalue $D_0(u)$ in 
(\ref{fac1})
which has all $\mu_k=\bar{\mu}_k=1$. This ratio can be computed
using the Euler-Maclaurin formula as in Appendix~B.
Using the expressions for the finitized Virasoro characters, we 
finally obtain the
{\it finitized} partition function for the Ising model on the Klein bottle
\begin{eqnarray}
Z^\sKlein(L;q) &=& X_0(\lfloor\case{L}{2}\rfloor;q^2) + |q| \,
X_{1/2}(\lfloor\case{L}{2}\rfloor;q^2) + |q|^{1/8} \,
X_{1/16}(\lfloor\case{L-1}{2}\rfloor;q^2)
\label{kbpf}
\end{eqnarray}

Notice that if we set $q=1$, then $Z(L;1)=r/2$ as given by
\eqref{effdim} corresponding to the counting of the number of real eigenvalues
left after cancellation of the complex conjugate pairs. As for the torus, the
even sector of the spectra with $R=1$ is given by $X_0$ and $X_{1/2}$ 
and the odd
sector with $R=-1$ is given by $X_{1/16}$.

\nsection{Ising Model on the M\"obius Strip}

In this section we consider the Ising model on the M\"obius strip.
Consider a square lattice consisting of a strip of $N$ columns and 
$M$ rows. Along
the vertical direction we apply a flip operator $\F$ before joining
row~$M$ to row~1. We apply integrable boundary conditions~\cite{BPO} 
on the left
and right edges of the strip. These boundary conditions labelled by 
$+$, $-$, $F$
(free) are specified by boundary triangle Boltzmann weights on the 
left and right
edges of the double row transfer matrix $\D(u)$. We build up the 
M\"obius strip by
$M$ applications of the double row transfer matrix. For compatibility 
with the flip
the boundary conditions on the left and right must be the same. This is because
the M\"obius strip only has one edge. By the spin reversal symmetry, 
the partition
function with $+$ boundary conditions is the same as the partition 
function with
$-$ boundary conditions so we do not need to consider $-$ boundary conditions.

The left and right
boundary weights are
\begin{equation}
\setlength{\unitlength}{13mm}
\text{1.6}{1.6}{0.8}{0.8}{}{\BLt{a}{b}{a}{u}}
\text{1}{1.6}{0.5}{0.8}{}{=}\lefttri{a}{b}{a}{u}
\text{1.2}{1.6}{.6}{0.8}{}{=}
\text{1.6}{1.6}{0.8}{0.8}{}{\BRt{a}{b}{a}{u}}
\text{1}{1.6}{0.5}{0.8}{}{=}\righttri{a}{b}{a}{u}
\end{equation}
where it is understood that these weights vanish if the adjacency 
condition is not
satisfied along the edges $(a,b)$. These weights must satisfy the boundary
Yang-Baxter equation.
For the Ising model,
we study the boundary conditions specified by the following non-zero weights
\begin{itemize}
\item{\textit{Fixed} $+$\,:} The edge spins are fixed to $a$=1
\begin{equation}
\BLt 121 u =\BRt 121 u = 1
\end{equation}
\item {\textit{Free} $F$\,:} The edge spins are fixed to $a$=2
\begin{equation}
\BLt 2 c 2 u =\BRt 2 c 2 u=\frac{1}{\sqrt2},\qquad c=1,3.
\end{equation}
\end{itemize}

For a strip with $N$ columns, the double row transfer matrix $\D(u)$
with rows  $\a=\{a_0,\dots,a_{N}\}$ and $\b=\{b_0,\dots,b_{N}\}$ is defined
diagrammatically~\cite{BPO} by
\setlength{\unitlength}{12mm}
\beq
\D(u)_{\smb a,\smb b}\ \;
=\raisebox{-1.4\unitlength}[1.6\unitlength][
1.4\unitlength]{\begin{picture}(4.5,3)
\multiput(0.5,0.5)(6,0){2}{\line(0,1){2}}
\multiput(1,0.5)(1,0){3}{\line(0,1){2}}
\multiput(5,0.5)(1,0){2}{\line(0,1){2}}
\multiput(1,0.5)(0,1){3}{\line(1,0){5}}
\put(1,1.5){\line(-1,2){0.5}}\put(1,1.5){\line(-1,-2){0.5}}
\put(6,1.5){\line(1,2){0.5}}\put(6,1.5){\line(1,-2){0.5}}
\put(0.5,0.45){\spos{t}{a_0}}\put(1,0.45){\spos{t}{a_0}}
\put(2,0.45){\spos{t}{a_1}}\put(3,0.45){\spos{t}{a_2}}
\put(5,0.45){\spos{t}{a_{N-1}}}\put(6,0.45){\spos{t}{a_{N}}}
\put(6.7,0.45){\spos{t}{a_{N}}}
\put(0.5,2.6){\spos{b}{b_0}}\put(1,2.6){\spos{b}{b_0}}
\put(2,2.6){\spos{b}{b_1}}\put(3,2.6){\spos{b}{b_2}}
\put(5,2.6){\spos{b}{b_{N-1}}}\put(6,2.6){\spos{b}{b_{N}}}
\put(6.7,2.6){\spos{b}{b_{N}}}
\put(1.05,1.45){\spos{tl}{c_0}}\put(2.05,1.45){\spos{tl}{c_1}}
\put(3.05,1.45){\spos{tl}{c_2}}\put(4.99,1.45){\spos{tr}{c_{N-1}}}
\put(5.99,1.45){\spos{tr}{c_{N}}}
\multiput(1.5,1)(1,0){2}{\spos{}{u}}\put(5.5,1){\spos{}{u}}
\multiput(1.5,2)(1,0){2}{\spos{}{\l\!-\!u}}
\put(5.5,2){\spos{}{\l\!-\!u}}
\put(0.71,1.5){\spos{}{\l\!-\!u}}\put(6.29,1.5){\spos{}{u}}
\multiput(0.5,0.5)(0,2){2}{\makebox(0.5,0){\dotfill}}
\multiput(6,0.5)(0,2){2}{\makebox(0.5,0){\dotfill}}
\multiput(1,1.5)(1,0){3}{\spos{}{\bullet}}
\multiput(5,1.5)(1,0){2}{\spos{}{\bullet}}
\multiput(1.1,0.62)(0,1){2}{\pp{}{\searrow}}
\multiput(2.1,0.62)(0,1){2}{\pp{}{\searrow}}
\multiput(5.1,0.62)(0,1){2}{\pp{}{\searrow}}

\end{picture}}\qquad\qquad
\qquad\qquad\eql{DRTM}
\eeq
For the Ising model on the M\"obius strip we require $a_0=b_0=a_N=b_N$.
The double row transfer matrices are real symmetric and crossing
symmetric~\cite{BPO}
\begin{equation}
\D(u)=\D^T(u)=\D(\lambda-u),\qquad u\in{\Bbb R}\label{Dsyms}.
\end{equation}
The dimension of the transfer matrices $\D(u)$ are given in terms of 
the adjacency
matrix $A$ by
\begin{equation}
\mbox{dim}\,\D(u)=[A^N]_{a_0,a_0}=
\cases{2^{L-1},&fixed $+$\cr 2^L,&free $\;F$}
\label{dim+F}
\end{equation}
where $N=2L$ must be even for both boundary conditions.

The flip operator $\F$ satisfying $\mbox{dim}\,\F=\mbox{dim}\,\D(u)$,
$\F^2=\I$, $\F^T=\F$ is defined by
\begin{equation}
\F_{\smb a,\smb b}=\prod_{j=0}^{N} \delta_{a_j,b_{N-j}}
\label{mbflipeqn}
\end{equation}
Proceeding as for the Klein Bottle, the number of
negative eigenvalues of $\F$ for the M\"obius strip is
\begin{equation}
2s_a=\left[A^N\right]_{a,a} - \sum_{k=1}^3
\left[A^{N/2}\right]_{a,k}
\label{numberflipmb}
\end{equation}
where the edge spin is $a=1$ for
fixed $+$ boundaries and $a=2$ for free $F$ boundaries.

The matrices $\D(u)$ commute with $\F$.
In fact, using \eqref{reflectweight} and (\ref{Dsyms}) we find
\begin{equation}
\F\D(u)\F=\D^T(\lambda-u)=\D^T(u)=\D(u).\label{FDF}
\end{equation}
The partition function is thus
\begin{equation}
Z^\sMobius_{MN}(u)=\mbox{Tr}\left[\F\D(u)^{M/2}\right]
=\sum_n F_n D_n(u)^{M/2}
\label{mbptfn}
\end{equation}
where the sum is over all eigenvalues.

More generally, for other lattice models with integrable boundary
conditions that are non-diagonal, the condition $a_0=b_0=a_N=b_N$ is 
not satisfied
and the last equality in (\ref{FDF}) does not hold.
In such cases it can be shown~\cite{BP2} that $\D(u)$ is similar to a
symmetric matrix $\tD(u)$ that does commute with $\F$.
Let us define the diagonal matrices
\begin{equation}
\cA_{\smb a,\smb b}=\left(S_{a_N}\over S_{a_0}\right)^{1/4}\prod_{j=0}^N
\delta_{a_j,b_j},\qquad
\cA^{-1}_{\smb a,\smb b}=\left(S_{a_0}\over S_{a_N}\right)^{1/4}\prod_{j=0}^N
\delta_{a_j,b_j}
\end{equation}
so that
\begin{equation}
\F\cA\F=\cA^{-1},\qquad \cA\F=\F\cA^{-1},\qquad \F\cA=\cA^{-1}\F. \label{FAF}
\end{equation}
Let us also define
\begin{equation}
\td \D(u)=\cA \D(u) \cA^{-1}.
\end{equation}
Then,
by using \eqref{reflectweight} we find
\begin{equation}
\td\D^T\!(u)=\td\D(u)\label{symtD}
\end{equation}
so that $\tD(u)$ is real symmetric for $u\in{\Bbb R}$.
It then also follows using the valid part of \eqref{FDF}, \eqref{FAF} and
\eqref{symtD} that $\F$ and $\tD(u)$ commute
\begin{eqnarray}
\F\tD(u)\F&=&\F\cA\D(u)\cA^{-1}\F=
\cA^{-1}\F\D(u)\F\cA\\
&=&\cA^{-1}\D^T\!(u)\cA
=(\cA\D(u)\cA^{-1})^T=\tD^T\!(u)=\tD(u)\nonumber
\end{eqnarray}
In this case $\D(u)$ should be replaced by $\tD(u)$ in the partition function
\eqref{mbptfn}.

\subsection{Finitized Ising partition function on the M\"obius strip}

The transfer matrices $\D(u)$
for both fixed and free boundary conditions form~\cite{BPO} a
commuting family $\D(u)\D(v)=\D(v)\D(u)$ whose eigenvalues $D(u)$
satisfy~\cite{OPW} the functional equation
\begin{equation}
D(u)D(u+\l)={\cos^{2(N+1)}2u-\sin^{2(N+1)}2u\over\cos 4u}
\end{equation}
where $\l=\pi/4$ is the crossing parameter.
For free boundary conditions the solution of this functional equation yields
\begin{equation}
D(u)=2^{-L}\prod_{k=1}^L\left(\cosec\left(\frac{\pi(2k-1)}{2(2L+1)}\right)+
\mu_k\sin 4u\right) \label{mbeigen}
\end{equation}
where $N=2L$ and $\mu_k=\pm1$ so that there are $2^L$ eigenvalues.
The same solution applies for fixed $+$ boundary conditions but with 
the constraint
\begin{equation}
\prod_{k=1}^L\mu_k=1.
\end{equation}
In this case there are $2^{L-1}$ eigenvalues in agreement with
(\ref{dim+F}).
The eigenvalues $D(u)$ are thus specified
by the set $\{\mu_k\}$.

The asymptotic behaviour of the finite-size partition function $Z_{NM}$ of the
critical Ising model on the M\"obius strip in the
limit of large $N$ and $M$ with the aspect ratio $M/N$ fixed is given by
\begin{equation}
Z^\sMobius_{NM}(u)=\mbox{Tr} [\F\D(u)^{M/2}]\sim\exp\left[-NM
\fb(u)-M\fs(u)\right]\, Z^\sMobius(q) \label{ZNMM}
\end{equation}
where $\D(u)$ is the double row transfer matrix, $\fb(u)$ is the bulk 
free energy,
$\fs(u)$ is the boundary free energy and $Z^\sMobius(q)$ is the 
universal conformal
partition function.
The leading corrections to each transfer matrix eigenvalue
$D(u)$ are of the form
\begin{equation}
\frac{1}{2} \log D(u) = -N\fb(u)-\fs(u)+\frac{\pi}{N}
\left(\frac{c}{24}-\Delta-k\right)\sin 4u+
\mbox{o}\left(\frac{1}{N}\right) \label{scal2}
\end{equation}
with $\Delta=0,1/2,1/16$ and $k$ a non-negative
integer. On the M\"obius strip we define
\begin{equation}
q = \exp\!\Big(\!-\pi\,{M\over N} \sin 4u\Big)
=|q_\sTorus|^{1/2}\label{modpar}
\end{equation}
where $q_\sTorus$ is the modular paramter on the torus (\ref{modtor}).
The {\it finitized} conformal partition function is thus
\begin{equation}
Z^\sMobius(L;q)=\sum_n F_n
\left(\frac{D_n}{D_{0}}\right)^{M/2}
  \label{confmb}
\end{equation}
where $n=0,1,\ldots$ labels the eigenvalues $D_n$ of $\D(u)$ and $F_n=\pm 1$ of
$\F$ for
$N$ columns and $D_{0}$ is the largest eigenvalue with all $\mu_k=1$.

Let $m$
be the number of $\mu_k$ such that $\mu_k=-1$. Assuming that $m$ remains finite
as $N\to\infty$, we see that
\begin{equation}
\lim_{N\rightarrow\infty} N\log\left(\frac{D(u)}{D_0(u)}\right)
=-\pi \sin 4u \sum_{k=1}^L (2k-1)\delta_{\mu_k,-1}
\end{equation}
Hence
\begin{equation}
\left(\frac{D(u)}{D_0(u)}\right)^{M/2}\simeq\;\prod_{k=1}^L
q^{(k-1/2)\,\delta_{\mu_k,-1}}
\end{equation}
where $q$ is given by \eqref{modpar}.
So the finitized partition function is
\begin{equation}
Z^\sMobius(L;q)=\left\{
\begin{array}{ll}
\sum_{\{\mu\}_L^+} F \displaystyle\prod_{k=1}^L
q^{(k-1/2)\,\delta_{\mu_k,-1}}
&\qquad\mbox{Fixed $+$}
\label{confmb2fix} \\[10pt]
\sum_{\{\mu\}_L} F \displaystyle\prod_{k=1}^L
q^{(k-1/2)\,\delta_{\mu_k,-1}}
&\qquad\mbox{Free $F$}
\end{array}
\right.
\end{equation}
where $\{\mu\}_L^{\pm}$ denotes the sequences restricted by
$\prod_{k=1}^L\mu_k=\pm1$.

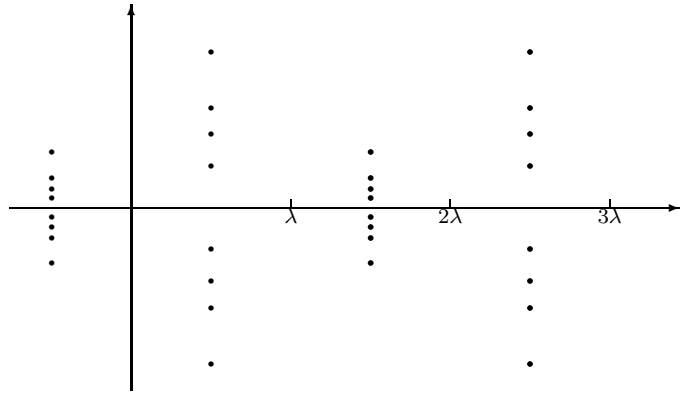
\begin{figure}[htb]
\setlength{\unitlength}{27mm}
\begin{center}
\begin{picture}(4.7,2.3)(-1.2,-1.4)
\multiput(1.1781,0.273282)(1.5708,0){1}{\md}
\multiput(1.1781,-0.273282)(1.5708,0){1}\md
\multiput(1.1781,0.146299)(1.5708,0){1}\md
\multiput(1.1781,-0.146299)(1.5708,0){1}\md
\multiput(1.1781,0.0945783)(1.5708,0){1}\md
\multiput(1.1781,-0.0945783)(1.5708,0){1}\md
\multiput(1.1781,0.0464651)(1.5708,0){1}\md
\multiput(1.1781,-0.0464651)(1.5708,0){1}\md
\multiput(1.1781,0.273282)(-1.5708,0){2}{\md}
\multiput(1.1781,-0.273282)(-1.5708,0){2}\md
\multiput(1.1781,0.146299)(-1.5708,0){2}\md
\multiput(1.1781,-0.146299)(-1.5708,0){2}\md
\multiput(1.1781,0.0945783)(-1.5708,0){2}\md
\multiput(1.1781,-0.0945783)(-1.5708,0){2}\md
\multiput(1.1781,0.0464651)(-1.5708,0){2}\md
\multiput(1.1781,-0.0464651)(-1.5708,0){2}\md
\multiput(1.9635,0.768516)(1.5708,0){1}\md
\multiput(1.9635,-0.768516)(1.5708,0){1}\md
\multiput(1.9635,0.492433)(1.5708,0){1}\md
\multiput(1.9635,-0.492433)(1.5708,0){1}\md
\multiput(1.9635,0.361832)(1.5708,0){1}\md
\multiput(1.9635,-0.361832)(1.5708,0){1}\md
\multiput(1.9635,0.20437)(1.5708,0){1}\md
\multiput(1.9635,-0.20437)(1.5708,0){1}\md
\multiput(1.9635,0.768516)(-1.5708,0){2}\md
\multiput(1.9635,-0.768516)(-1.5708,0){2}\md
\multiput(1.9635,0.492433)(-1.5708,0){2}\md
\multiput(1.9635,-0.492433)(-1.5708,0){2}\md
\multiput(1.9635,0.361832)(-1.5708,0){2}\md
\multiput(1.9635,-0.361832)(-1.5708,0){2}\md
\multiput(1.9635,0.20437)(-1.5708,0){2}\md
\multiput(1.9635,-0.20437)(-1.5708,0){2}\md
\put(0,0){\vector(1,0){2.7}}
\put(0,0){\line(-1,0){0.6}}
\put(0.785398,0){\spos{t}{\lambda}}
\put(1.5708,0){\spos{t}{2\lambda}}
\put(2.35619,0){\spos{t}{3\lambda}}
\put(0.785398,0){\line(0,1){0.04}}
\put(1.5708,0){\line(0,1){0.04}}
\put(2.35619,0){\line(0,1){0.04}}
\put(0,0){\vector(0,1){1}}
\put(0,0){\line(0,-1){0.9}}
\end{picture}
\end{center}
\vspace{-1.7cm}
\caption{Zeros of a typical eigenvalue of $\D(u)$ in the period strip
$-\lambda<\Re(u)\le 3\lambda$ in the complex $u$-plane. The relevant 
analyticity
strip is $-\lambda/2\le \Re(u)\le 3\lambda/2$.}
\label{mbzeroplot}
\end{figure}

 From \cite{OPW}, the eigenvalues
\eqref{mbeigen} can be written as
\begin{equation}
D(u)=\epsilon\left(4ie^{4iu}\right)^{-L}\prod^{2L+1}_{
k=1 \atop
k\neq L+1}
\left(e^{4iu}+i\mu_k\tan\Big(\frac{\pi(2k-1)}{4(2L+1)}\Big)
\right) \label{mbeigenraw}
\end{equation}
with $\epsilon=\prod_{k=1}^L\mu_k$.
As in Section \ref{finiteklein},
these eigenvalues are determined by their zeros in the complex $u$-plane as
given by \eqref{eigenexpress}.
Since the zeros occur in complex conjugate pairs we will only
look at the zeros of $D(u)$ in the upper-half plane given by
\begin{eqnarray}
u_k&=&\frac{\pi}{8}\mub_k-\frac{\pi}{4}
-\frac{i}{4}\log t_{L+1-k}\quad \mbox{mod $\displaystyle{\pi\over 2}$},\qquad
k=1,2,\ldots,L \label{mbzero}
\end{eqnarray}
where $\mub_k=\mu_{L+1-k}$ and
$t_k$\,=$\,\tan\left(\frac{\pi(2k-1)}{4(2L+1)}
\right)$.
The zeros occur on certain fixed lines and are
classified into 1-strings and 2-strings. The relevant analyticity strip is
$-\pi/8\le \Re(u)\le 3\pi/8$ and a 1-string is  a single zero $u_k$ such that
\begin{equation}
\Re(u_k)=\frac{\pi}{8}
\end{equation}
A 2-string is a pair of zeros $(u_k,u_k')$ with equal imaginary part and
\begin{equation}
\left(\Re(u_k),\Re(u_k')\right)=\left(-\frac{\pi}{8}\,,\frac{3\pi}{8}\right)
\end{equation}
A distribution of zeros for a typical eigenvalue $D(u)$ for $L=8$
is depicted in Figure~\ref{mbzeroplot}.

The eigenvalues are completely classified by the string contents and 
the relative
orderings of the strings in the analyticity strip. For each eigenvalue $D(u)$,
there are $L=N/2$ distinct strings on the upper-half complex $u$-plane.
Let $m$ be the number of 1-strings in the analyticity strip
and $n$ the number of 2-strings. Then the string contents $m$ and $n$ 
satisfy the
simple $(m,n)$ system~\cite{Berkovich,Melzer}
\begin{equation}
m+n=L.
\end{equation}
Note that $j=1$ labels the 1- and 2-strings closest to the real axis and for
each eigenvalue the ordering of the imaginary parts of the strings is
\begin{equation}
\Im(u_L)>\Im (u_{L-1})>\ldots >\Im (u_2)>\Im (u_1)>0.
\end{equation}
  From \eqref{mbzero}, a string at $\Im(u_k)$ is a  1-string if $\mub_k=-1$
and a 2-string if $\mub_k=1$. Suppose that
\begin{equation}
\mub_k=-1,\qquad k=k_1,k_2,\ldots,k_m,\qquad k_1<k_2<\ldots<k_m.
\end{equation}
Then we define the quantum numbers
\begin{equation}
I_j=\sum_{k=k_j}^L{1\over 2}(1+\mub_k)
=\{\mbox{\# of 2-strings above given 1-string $k_j$}\}
\end{equation}
Alternatively, these can be defined recursively by
\begin{equation}
I_j-I_{j-1}+k_j-k_{j-1}=1,\qquad j=1,2,\ldots,m
\end{equation}
with $k_0=0$, $I_0=L-m$ and solution
\begin{equation}
I_j=L-m+j-k_j,\qquad j=0,1,\ldots,m.
\end{equation}
The quantum numbers $I=(I_1,I_2,\ldots,I_m)$ satisfy
\begin{equation}
L-m=n\ge I_1\ge I_2\ge\ldots\ge I_m\ge 0
\end{equation}
and uniquely label all of the eigenvalues with
given string content $m$.

The commuting matrices $\D(u)$ and $\F$ have a common set of 
eigenvectors. If the
eigenvector $\x$ associated with the eigenvalue $D(u)$ satisfies 
$\F\x=\x$ we say
that the eigenvalue $D(u)$ has parity $F=+1$ under the flip. Otherwise, if the
eigenvector $\x$ satisfies $\F\x=-\x$, we say that the eigenvalue 
$D(u)$ has parity
$F=-1$ under the flip. For both fixed and free boundary conditions, we find the
following results from numerics:
\begin{enumerate}
\item An eigenvalue $D(u)$ with quantum numbers
$(I_1,I_2,\ldots,I_m)=(0,0,\ldots,0)$, such that the 1-strings are all further
from the real axis than the 2-strings, has parity
$F=+1$ under the flip
$\F$.
\item An eigenvalue $D(u)$ with quantum numbers
$(I_1,I_2,\ldots,I_m)$ has parity
\begin{equation}
F=(-1)^{I_1+I_2+\ldots+I_m}
\end{equation}
under the flip $\F$. This implies that interchanging the position of a 1-string
with the position of an adjacent 2-string changes the sign of the parity.
\end{enumerate}
These parity properties can not be obtained by studying the eigenvalues alone.
Rather these parity properties reflect deep properties of the 
eigenvectors which
we do not study here. Nevertheless, these observations allow the flip 
eigenvalue
$F=\pm1$ to be read off directly from the eigenvalue $D(u)$ just by 
looking at the
locations of its zeros in the complex $u$-plane.

For fixed $+$ boundaries on the M\"obius strip, it now follows from
\eqref{confmb2fix} and our numerical observations that
\begin{eqnarray}
Z^+(L;q)&=&\sum_{\{\mu\}_L^+} F
\prod_{k=1}^L q^{(k-1/2)\,\delta_{\mu_k,-1}}
=\sum_{m\;{\rm even}} q^{m^2\over 2}\sum_I
(-q)^{\sum_{j=1}^m I_j}\nonumber\\
&=&\sum_{m\;{\rm even}} q^{m^2\over 2}
\Mult{L}{m}_{\!(-q)}
= X_0(L;-q) \label{confmb3}
\end{eqnarray}
where we have used (\ref{Gaussianpoly}) and the facts that $m$ is even,
$\mu_k=\mub_{L+1-k}$ and
\begin{equation}
\sum_{k=1}^L \Big(k-{1\over 2}\Big)\,\delta_{\mu_k,-1}
=\sum_{j=1}^m \Big(L+1-k_j-{1\over 2}\Big)={m^2\over 2}+\sum_{j=1}^m I_j.
\end{equation}
Similarly, for free boundaries on the M\"obius strip,
the finitized partition function is
\begin{eqnarray}
Z^F(L;q)&\!\!=\!\!&\sum_{\{\mu\}_L^+}F\prod_{k=1}^L
q^{(k-1/2)\,\delta_{\mu_k,-1}}
+ \sum_{\{\mu\}_L^-} F\prod_{k=1}^L
q^{(k-1/2)\,\delta_{\mu_k,-1}}\nonumber\\
&\!\!=\!\!&
\sum_{m\;{\rm even}} q^{m^2\over 2}\sum_I(-q)^{\sum_{j=1}^m I_j}
+q^{1/2}\sum_{m\;{\rm odd}} q^{m^2-1\over 2}\sum_I(-q)^{\sum_{j=1}^m I_j}
\nonumber\\
&\!\!=\!\!& X_0(L;-q)+q^{1/2}X_{1/2}(L;-q)\label{confmbfree}
\end{eqnarray}
If we let $L\rightarrow\infty$ we obtain
the conformal partition functions of the Ising model on the M\"obius 
strip in terms
of Virasoro characters. Notice also that removing the flip $F$ in 
(\ref{confmb3})
and (\ref{confmbfree}) gives the finitized cylinder partition functions
$Z^{+|+}_{NM}(q)$ and
$Z^{F|F}_{NM}(q)$ respectively. For the M\"obius strip with free boundary
conditions, the even sector of the spectra with $R=1$ is given by $X_0$ and the
odd sector with $R=-1$ is given by $X_{1/2}$.

Lastly we note that, in agreement with \eqref{numberflipmb}, the 
number of negative
eigenvalues of $\F$ is
\begin{equation}
\lim_{q\to 1}{1\over 2}\Big[Z^{a|a}(L;q)-Z^a(L;q)\Big]
=\cases{\frac{1}{2}\left(
2^{L-1}-2^{\lfloor\frac{L}{2}\rfloor}\right),&$a=1$ or $+$\cr
\frac{1}{2}\left(2^L-2^{\lfloor\frac{L+1}{2}\rfloor}\right)\rule{0in}{.25in},
&$a=2$ or $F$
}
\end{equation}
where we have used
\begin{equation}
{\textstyle\frac{1}{2}\Big(\Mult{L}{\os}_{q=1}-
\Mult{L}{\os}_{q=-1}\Big)}
=\left\{\begin{array}{ll}
\frac{1}{2} {L\choose \os}, & \mbox{\qquad $L$ odd, $\os$
even}
\\
\frac{1}{2}\Big[
{L\choose \os} -
{{\lfloor
\frac{L}{2}\rfloor}\choose{\lfloor\frac{\os}{2}\rfloor}}
\Big]\rule{0in}{.25in}, &\mbox{\qquad otherwise.}
\end{array}
\right.
\end{equation}

\nsection{Summary and Discussion}\label{summary}
In this paper we have obtained the finitized conformal partitions of the Ising
model on the Klein bottle and M\"obius strip using Yang-Baxter 
techniques and the
solution of functional equations. Changing the topology from the 
torus to the Klein
bottle or the cylinder to the M\"obius strip is achieved by inserting a flip
operator before gluing the boundaries with the trace. Under the 
action of the flip
operator, a parity $F=\pm1$ is assigned to the eigenvalues of the commuting
transfer matrices. The task is thus reduced to finding the parities of the
eigenvalues of the relevant transfer matrices. Although in this paper 
we have only
considered the Ising model, our methods and observations apply more 
generally to
solvable lattice models as we now explain.

For the Klein bottle, it was shown analytically that all the complex 
eigenvalues
of the single row transfer matrix cancel with their complex
conjugates. The partition function for Ising model on the Klein bottle
is thus given by a projection from the partition function on the torus
\begin{eqnarray}
Z^{\sTorus}(q)&=&|\chi_0(q)|^2+|\chi_{1/2}(q)|^2+|\chi_{1/16}(q)|^2
\nonumber\\
&\mapsto&\chi_0(q^2)+\chi_{1/2}(q^2)+\chi_{1/16}(q^2)
\;=\;Z^{\sKlein}(q)
\label{finalKlein}
\end{eqnarray}
where only the left-right chiral symmetric part is retained.
In fact, from numerics, this projection is valid for any
$A_L$ model with $L\geq3$ so that we observe the projection
\begin{equation}
Z^{\sTorus}_{A_L}(q)=\sum_{\Delta} |\chi_{\Delta}(q)|^2
\mapsto\sum_{\Delta} \chi_{\Delta}(|q|^2)=Z_{A_L}^{\sKlein}(q)
\end{equation}
where the sum is over the Kac table.
Similarly, we have checked that the same mechanism works for the $D_4$ or Potts
model
\begin{eqnarray}
Z^{\sTorus}_{D_4}(q)&=&|\chi_0(q)+\chi_3(q)|^2+
|\chi_{2/5}(q)+\chi_{7/5}(q)|^2+2|\chi_{1/15}(q)|^2+2|\chi_{2/3}(q)|^2
\nonumber\\
&\mapsto&\chi_0(q^2)+\chi_3(q^2)+\chi_{2/5}(q^2)+\chi_{7/5}(q^2)
+2\chi_{1/15}(q^2)+2\chi_{2/3}(q^2)\\
&=&Z^{\sKlein}_{D_4}(q)\nonumber
\end{eqnarray}

For the M\"obius strip, the modular parameter effectively changes from $q$ to
$-q$ under the action of the flip operator which transforms
the topology from the cylinder to the M\"obius strip. More specifically, since
$q=\mbox{exp}\,(2\pi\i\tau)$,
this is equivalent to saying that $\tau\mapsto \tau+1/2$. Again we find
numerically that this projection mechanism applies to other $A_L$ 
models such as,
for example, $A_4$ with the $(r,s)=(2,2)$ boundary condition
\begin{eqnarray}
Z^{\sCyl,\;(2,2)}_{A_4}(q)&\!\!=\!\!&\chi_{1,1}(q)+\chi_{1,3}(q)+
\chi_{3,1}(q)+\chi_{3,3}(q)\nonumber\\
&\!\!\mapsto\!\!& \tilde\chi_{1,1}(-q)+\tilde\chi_{1,3}(-q)+
\tilde\chi_{3,1}(-q)+\tilde\chi_{3,3}(-q)\\
&=&Z_{A_4}^{\sMobius,\;(2,2)}(q)\nonumber
\end{eqnarray}
where the tilde on the Virasoro characters means that the leading 
fractional power
of $q$ is replaced by the same fractional power of $|q|$.

Our results for the Ising model are in agreement with \cite{Yamaguchi,LuWu2}.
To compare the results with \cite{Yamaguchi}, we use the identities
\begin{equation}
\sqrt{\frac{\theta_3}{\eta}}=q^{-1/48}\prod_{n=0}^\infty (1+q^{n+1/2}),
\qquad\qquad
\sqrt{\frac{\theta_4}{\eta}}=q^{-1/48}\prod_{n=0}^\infty (1-q^{n+1/2})
\end{equation} 
and $\tau=i\frac{R}{4L}+1/2$
to re-express $Z_{\mbox{\scriptsize fixed($\pm$)}}(\tau)$ and 
$Z^{(1)}_{\mbox{\scriptsize free}}(\tau)$ in (11,12) of \cite {Yamaguchi}
from theta and eta functions to q-product representation (Note that
$Z^{(2)}_{\mbox{\scriptsize free}}(\tau)$ corresponds to anti-periodic boundary which we don't consider here). 

In \cite{LuWu2}, again we first re-express the theta functions into q-product representaion. For Klein bottle, we use the fact that 
$\prod_{n=1}^N(1-q^{2n-1})(1+q^n)=1+O(q^{N+1})$, and substituting the aspect ratio $\xi$ into $q$, we obtain the same exact formula \eqref{finalKlein} in terms of q-products (\ref{chi0},\ref{chi12},\ref{chi16}). For M\"obius strip, we confirm the result by  series expansion about $q=0$ up to 100 terms using {\it Mathematica}. 

Finally, we remark that it would be of interest to generalize our 
results on the
Klein bottle and M\"obius strip to the more general twisted boundary conditions
discussed recently in the context of conformal field theory by Petkova and
Zuber~\cite{PetZuber}.

\section*{Appendix~A: Some Linear Algebra}
\rnc{\theequation}{A.\arabic{equation}}
\setcounter{equation}{0}

We list some useful properties of $\T(u)$ and $\F$:
\begin{enumerate}
\item
If $\T(u)\x=T(u)\x$ and $\T(u)$ is real then
\begin{equation}
\T(u)\overline{\x}=\overline{\T(u)\x}=\overline{T(u)}\overline{\x}
\label{complexvector}
\end{equation}
so that if $\x$ is real then $\overline{T(u)}=T(u)$ and $T(u)$ is real.
\item If $\T(u)$ is a real normal matrix with $\T^T(u)=\T(\lambda-u)$ then the
eigenvalue of $\T^T(u)=\T(\lambda-u)$ is $\overline{T(u)}$.
This follows since, if $\x$ is a common
eigenvector of $\T(u)$ and $\T(\lambda-u)$ so that $\T(u)\x=T(u)\x$
and $\T(\lambda-u)\x=T(\lambda-u)\x$, then
\begin{eqnarray}
T(\lambda-u)|\x|^2=\x^\dagger \T(\lambda-u)\x=
\x^\dagger \T^T(u)\x=(\T(u)\overline{\x})^T\x
=\overline{T(u)}|\x|^2\label{org1}
\end{eqnarray}
and so $T(\lambda-u)=\overline{T(u)}$.
\item Suppose further that $T_i$ and $T_j$ are eigenvalues of $\T=\T(u_0)$ and
$T_i\neq T_j$, then the corresponding eigenvectors
$\x_i$ and $\x_j$ are orthogonal. This follows since
\begin{eqnarray}
\x_i^\dagger(\T\x_j)=T_j\x_i^\dagger\x_j =
\big(\T^T\x_i\big)^\dagger\x_j=\big(\overline{T_i}\,\x_i\big)^\dag
\x_j  =T_i\,\x_i^\dagger\x_j
\end{eqnarray}
which implies $(T_i-T_j)\x_i^\dagger\x_j=0$ and $\x_i^\dagger\x_j=0$.
\item For any eigenvector $\x$ of $\T(u)$, $\F\x$ is also an
eigenvector of $\T(u)$ with the corresponding eigenvalue
$\overline{T(u)}$:
\begin{equation}
\T(u)\left(\F\x\right)=(\T(u)\F)\x=\left(\F\T(\lambda-u)\right)\x
      =\overline{T(u)}\F\x\label{cconj2}
\end{equation}
\item By \eqref{Fx} and the normality of $\T(u)$, $\F$ and
$\D(u)=\T(\lambda-u)\T(u)$ commute:
\begin{eqnarray}
\mbox{}\hspace{-.2in}
\F\big(\T(\lambda-u)\T(u)\big)=\T(u)\F\T(u)=\T(u)\T(\lambda-u)\F
=\big(\T(\lambda-u)\T(u)\big)\F \label{commuteFD}
\end{eqnarray}
\end{enumerate}

\section*{Appendix~B: Euler Maclaurin}
\rnc{\theequation}{B.\arabic{equation}}
\setcounter{equation}{0}

For the single row transfer matrix $\T(u)$, we compute the scaling limit of the
largest eigenvalues in the $R=+1$ and $R=-1$ sectors by applying the
midpoint and endpoint Euler-Maclaurin formulas
\begin{eqnarray}
&&\mbox{}\hspace{-.25in}\sum_{k=1}^{m}f\Big(a+(2k-1)h\Big)=\frac{1}{h}\int_a^b
f(t)\,dt-\frac{h}{24}
[f'(b)-f'(a)]+\mbox{O}\big(h^2\big)\\
&&\mbox{}\hspace{-.5in}\sum_{k=0}^{m}f(a+kh)=\frac{1}{h}\int_a^b
f(t)\,dt+{1\over 2}[f(b)+f(a)]+\frac{h}{12}
[f'(b)-f'(a)]+\mbox{O}\big(h^2\big)
\end{eqnarray}
where $b=a+mh$. These formulas are valid if $f(t)$ is twice 
differentiable on the
closed interval $[a,b]$.

In the $R=+1$ sector, the logarithm of the largest
eigenvalue can be written as
\begin{equation}
\log T_0(u)=\frac{1}{2}(1-L)\log2
+\frac{1}{2}\sum_{k=1}^{L}
\log\Big[\sin 4u +\csc\left(\textstyle\frac{\pi(2k-1)}{2L}
\right)\Big]
\end{equation}
Suppressing the dependence on $u$, we define the function
\begin{equation}
f(t)=\log\Big[t(\pi-t) (\sin 4u + \csc t)\Big]
\end{equation}
so that
\begin{equation}
\log T_0(u)=\frac{1}{2}(1-L)\log2
+\sum_{k=1}^{L} f\left(\textstyle\frac{\pi(2k-1)}{2L}
\right)-2\sum_{k=1}^{L}
\log\left(\textstyle\frac{\pi(2k-1)}{2L}
\right)
\end{equation}
We define $f$ in this way to remove the singularities of
$\csc t$ at $t=0,\pi$. Applying the midpoint Euler-Maclaurin formula and the
asymptotic expansion of the gamma function
\begin{equation}
\log \Gamma(z)=\big(z-{\textstyle\frac{1}{2}}\big)\log
z-z+{\textstyle\frac{1}{2}}\log{2\pi}+\frac{1}{12z}+\mbox{O}\big(z^{-2}\big)
\end{equation}
we obtain
\begin{equation}
\log T_0(u)=\frac{N}{4\pi}\int_0^{\pi}\log\frac{1}{2}\Big(
\sin4u+\csc t\Big)dt+\frac{\pi}{12N}\sin4u+\mbox{o}\left(
\frac{1}{N}
\right)
\end{equation}
which agrees with \eqref{pcorr} with
$c=1/2$, $\Delta=\overline{\Delta}=k=\bar{k}=0$
    and bulk free energy
\begin{equation}
\fb=-\frac{1}{4\pi}\int_0^{\pi}\log\frac{1}{2}\Big(
\sin4u+\csc t\Big)dt. \label{bulkfree}
\end{equation}

Similarly, using the endpoint Euler-Maclaurin formula in the $R=-1$ sector,
the logarithm of the largest eigenvalue is
\begin{eqnarray}
\mbox{}\hspace{-.4in}\log T_1(u)
&\!\!\!=\!\!\!&\frac{1}{2}\log L+\frac{1-L}{2} \log2
+\frac{1}{2}\sum_{k=1}^{L-1}\log\Big[\sin 4u +\csc\left(
\textstyle\frac{\pi k}{L}\right)\Big]\nonumber\\
&\!\!\!=\!\!\!&\frac{N}{4\pi}\int_0^{\pi}\log\frac{1}{2}\Big(
\sin4u+\csc t\Big)\,dt-\frac{\pi}{6N}\sin4u+\mbox{o}\left(
\frac{1}{N}\right)
\end{eqnarray}
with $\fb$ defined by \eqref{bulkfree} and $\Delta=1/16$.
Hence we conclude that
\begin{equation}
\log\big[{T_1(u)}/{T_0(u)}\big]=-\frac{\pi}{4N}\sin 4u
+\mbox{o}\left(\frac{1}{N}\right)
\end{equation}
and we obtain the factor $|q|^{1/8}$ in \eqref{kbpf}.

\subsubsection*{Acknowledgements}
This work is supported by the Australian Research Council. We thank
Christian Mercat and Vladimir Rittenberg for useful discussions and a critical
reading of the manuscript. We also thank the referee for helpful 
comments and advice.

\end{document}